\documentclass[
 reprint,
 superscriptaddress,
 amsmath,
 amssymb,
 aps,
 nofootinbib
]{revtex4-2}

\usepackage[utf8]{inputenc}
\usepackage[T1]{fontenc}
\usepackage{graphicx}
\usepackage{dcolumn}
\usepackage{bm}
\usepackage{natbib}
\DeclareUnicodeCharacter{2299}{$_\odot$}
\usepackage{marvosym}
\usepackage[normalem]{ulem}
\usepackage{multirow}

\usepackage[usenames]{color}
\usepackage[dvipsnames]{xcolor}

\usepackage{hyperref}
\hypersetup{colorlinks,linkcolor={black},citecolor={blue},urlcolor={blue}}

\usepackage[detect-all]{siunitx}

\begin{document}

\title{Gamma-ray emission from decays of boosted nuclei in protomagnetar jets}

\author{Sean Heston}
\email{seanh125@vt.edu}
\affiliation{Center for Neutrino Physics, Department of Physics, Virginia Tech, Blacksburg, Virginia 24061, USA}

\author{Nick Ekanger}
\email{ekangernj@astr.tohoku.ac.jp}
\affiliation{Frontier Research Institute for Interdisciplinary Sciences, Tohoku University, Sendai 980-8578, Japan}
\affiliation{Astronomical Institute, Graduate School of Science, Tohoku University, Sendai 980-8578, Japan}
\affiliation{Center for Neutrino Physics, Department of Physics, Virginia Tech, Blacksburg, Virginia 24061, USA}

\author{Shunsaku Horiuchi}
\email{horiuchi@vt.edu}
\affiliation{Department of Physics, Institute of Science Tokyo, 2-12-1 Ookayama, Meguro-ku,
Tokyo 152-8551, Japan}
\affiliation{Center for Neutrino Physics, Department of Physics, Virginia Tech, Blacksburg, Virginia 24061, USA}
\affiliation{Kavli IPMU (WPI), UTIAS, The University of Tokyo, Kashiwa, Chiba 277-8583, Japan}

\date{\today}

\begin{abstract}
{We examine the detectability of $\gamma$-ray emission originating from the radioactive decays of unstable nuclei that are synthesized in relativistic outflows launched in magnetorotational core-collapse supernovae. The observed lines have enhanced energies due to the Lorentz boosted nuclei and can also be seen until later times due to time dilation of the rest-frame half-lives. We find that instruments like \textit{e-ASTROGAM} and \textit{INTEGRAL/SPI} are sensitive to these boosted line emissions from hundreds of keV to tens of MeV at a distance of 10 kpc over timescales of tens of days. For favorable viewing angles, these decays can be detected to extragalactic distances for rapidly spinning protomagnetar models. On the other hand, detection for off-axis jets is challenging, even for a supernova at the Galactic Center. Measuring multiple decay lines in addition to the integrated luminosity over $\sim10$\:days postbounce would allow for the ability to distinguish between models and shed light on central engine properties like magnetic field and spin.}
\end{abstract}

\maketitle
\section{Introduction}\label{Intro}

The ability to determine the central engines that power the explosions of core-collapse supernovae (CCSNe) is important for testing current models and understanding the variety of astronomical transients discovered. One potential central engine is a highly magnetized and rapidly rotating protoneutron star, also known as a protomagnetar (PM) \cite{1992ApJ...392L...9D,1992Natur.357..472U}. These magnetorotational energy reservoirs may power transients like gamma-ray bursts (GRBs), super-luminous supernovae, hypernovae, and other extreme transients  \cite{Zhang:2000wx,Margutti:2013pra,Lu:2014oda,Margutti:2014gha,Kashiyama:2015eua}. These engines may also drive signatures of high-energy physics like ultrahigh energy cosmic rays \cite{Arons:2002yj,Kotera_2011_uhecr}, high-energy neutrinos \cite{Murase:2010gj,Murase:2013mpa,Murase:2017pfe,Bhattacharya:2022btx,Carpio:2023wyt}, and $r$-process nucleosynthesis of nuclei heavier than the iron group (see recent simulations \cite{Nishimura:2015nca,Halevi:2018vgp,Mosta:2017geb,Reichert:2024vyd}).

The proposed supernova explosions from magnetorotational central engines differ from those of typical CCSNe that are thought to be neutrino driven. In particular, magnetorotational central engines are likely to give rise to unique features beyond the standard production of $\mathcal{O}(10)$\:MeV CCSN neutrinos. One example is in the nucleosynthesis that occurs. The optical light curve that is normally associated with CCSNe is partially powered by the decay of radioactive nuclei within the ejecta, primarily of iron-group elements and lighter nuclei \cite{1982ApJ...253..785A,Hamuy:2002qx,Sawada:2023iau}. However, with a higher neutron-to-proton ratio and a lower entropy per baryon, magnetorotational CCSNe may synthesize heavier elements \cite{Nishimura:2015nca,Halevi:2018vgp,Reichert:2024vyd} and shed light on how heavy nuclei are sourced in our universe. If heavy nuclei are synthesized by supernovalike transients, this can be tested observationally \cite{Anand:2023ujd,Blanchard:2023zls,Rastinejad:2023ipp} and through galactic chemical evolution models \cite{Cote:2017evr,Hotokezaka:2018aui,Kobayashi:2020jes}. Another feature is the presence of relativistic jets. In some models, PMs can power relativistic jets \cite{Qian_1996, Metzger2011a}, powered by the transfer of the rotational energy of the rapidly rotating and highly magnetized PM. This scenario is employed to explain the origin of long GRBs from CCSNe (see, e.g., Refs.~\cite{Metzger2011a,Margutti:2013pra,Lu:2014oda}). There has been an ongoing effort to identify the multimessenger signatures of such a jet using, e.g., neutrinos \cite{Derishev:1999grb,Meszaros:2001ms,Razzaque:2004yv,bahcall20005,MESZAROS2005307,Ando:2005xi,Horiuchi:2007xi,Murase:2008grb,Murase:2013ffa,He:2018lwb,Bhattacharya:2022btx,Carpio:2023wyt}, very-high energy photons \cite{Murase:2008grb,Murase&Beacom_2010,Bosnjak:2023vrm}, and/or ultrahigh energy cosmic rays \cite{MGH11,Zhang:2017moz}.

In this work, we consider another signature of PM driven jets. We examine nuclei that are synthesized within the jet, including radioactive species that decay with various half-lives. Since these jets can reach relativistic bulk velocities, both the energy of any nuclear decay lines and the half-lives of radioactive nuclei are boosted by the bulk Lorentz factor of the jet. We explore a range of PM parameters, and demonstrate that decay lines may escape and provide detectable $\gamma$-ray line features before the jet enters the afterglow phase. Furthermore, some nuclei may also be accelerated to nonthermal spectra, similar to scenarios where magnetorotational central engines are employed to explain GRBs. In this case, the decay lines from accelerated nuclei blend to a power-law spectrum reaching  significant energies. We show that both the decay lines and the power-law continuum are potentially detectable from nearby galaxies for favorable PM parameters and viewing angles. 

The paper is structured as follows. In Sec.~\ref{Protomag}, we describe our physical models and detail the interactions our jet will undergo with time. In Sec.~\ref{Heavy Nuclei}, we discuss the nuclei that are synthesized within our jets and how they may avoid disintegration back into nucleons. In Sec.~\ref{Signal}, we explore the decays from unstable nuclei that are synthesized and how they result in a $\gamma$-ray signal. In Sec.~\ref{Non-thermal Signal}, we consider the scenario where some nonthermal particle acceleration occurs and the effect this has on the resultant signal. Finally, we discuss our assumptions in Sec.~\ref{Discussion} and summarize in Sec.~\ref{Summary}.

\section{Protomagnetar jets}\label{Protomag}

We describe a toy model for the generation and launch of a relativistic jet from an initially neutrino-driven outflow powered by a PM formed during a stellar core collapse event. We discuss the physical properties of the jet and their dependencies on the PM. We then discuss whether nuclei decay lines can escape these jets or not. Finally, we discuss the timing of an afterglow which is likely to overwhelm the nuclei decay line signals. 

\subsection{Jet mechanism and model}\label{Jet Model}

We describe the model used for PM driven outflows. Throughout this work, we refer to the neutrino-driven mass loss as a ``wind,'' the collimated structure as a ``jet,'' and a general term ``outflow'' for mass loss structures. We model the wind and jet largely following Ref.~\cite{Ekanger:2022tia} (see also Refs.~\cite{Qian_1996,Metzger2011a,Bhattacharya:2021cjc}) and consider PMs described by an initial rotation period $P_i$ and dipole field strength $B_{\rm dip}$. Neutrinos drive a wind with the mass-loss rate given by \cite{Metzger2011a}
\begin{align}\label{Mass Loss Rate}
    \nonumber \Dot{M}_w&= (5\times10^{-5}~\textrm{M}_\odot~\textrm{s}^{-1})\left[\frac{L_{\nu}}{10^{52}\:{\rm erg~s^{-1}}}\left(\frac{\varepsilon_{\nu}}{10\:{\rm MeV}}\right)^2\right]^{5/3}\\
    &\times \mathcal{F}_{\rm mag}\left(C_{\rm inel}\frac{R_{\rm NS}}{10^6\:{\rm cm}}\right)^{5/3}\left(\frac{M_{\rm NS}}{1.4\:\textrm{M}_{\odot}}\right)^{-2},
\end{align}
where there are several $B_{\rm dip}$ and $P_i$ dependent correction factors. First, $\mathcal{F}_{\rm mag}=f_{\rm op}f_{\rm cen}$ is a correction factor that considers the fraction of the PM surface threaded by open magnetic field lines ($f_{\rm op}$) and an increase to the mass loss rate due to magnetocentrifugal slinging ($f_{\rm cent}$). There is also a correction factor to account for neutrino-electron inelastic scatterings ($C_{\rm es}$). The other factors depend on the progenitor and average neutrino properties. We follow Ref.~\cite{Metzger2011a} but with neutrino light curves, i.e., the evolution of the neutrino luminosity ($L_{\nu}$) and mean energy ($\varepsilon_{\nu}$), from Ref.~\cite{Pons1999}. These assume a PM mass ($M_{\rm NS}$) of $1.4$\:$\textrm{M}_{\odot}$ and a radius ($R_{\rm NS}$) of $10^6$\:cm. The entropy and expansion timescale of the wind can also be described in relation to these quantities, and are given by
\begin{align}\label{Entropy}
    \nonumber S&=(88.5\ {k_{\rm B}\ \rm nuc^{-1}})\ \left[\frac{L_{\nu}}{10^{52}\:{\rm erg~s^{-1}}}\left(\frac{\varepsilon_{\nu}}{10\:{\rm MeV}}\right)^2\right]^{-1/6}\\
    &\times C_{\rm inel}^{-1/6}\left(\frac{R_{\rm NS}}{10^6\:{\rm cm}}\right)^{-2/3}\left(\frac{M_{\rm NS}}{1.4\:\textrm{M}_{\odot}}\right),
\end{align}
and
\begin{align}\label{Expansion Timescale}
    \nonumber \tau_{\textrm{exp}}&=(68.4\ {\rm ms})\ \left[\frac{L_{\nu}}{10^{52}\:{\rm erg~s^{-1}}}\left(\frac{\varepsilon_{\nu}}{10\:{\rm MeV}}\right)^2\right]^{-1}\\
    &\times f_{\rm op}C_{\rm inel}^{-1}\left(\frac{R_{\rm NS}}{10^6\:{\rm cm}}\right)\left(\frac{M_{\rm NS}}{1.4\:\textrm{M}_{\odot}}\right).
\end{align}

These quantities are related to the density and temperature of the neutrino-driven wind, which are necessary for determining the composition as a result of nucleosynthesis. The density and temperature are given by
\begin{equation}\label{Density}
    \rho=\frac{\tau_{\textrm{exp}}\Dot{M}_\mathrm{w}}{4\pi r^3} \mathcal{F}_{\rm mag}^{-1},
\end{equation}
\begin{equation}\label{Temperature}
    T=\left(\frac{45\rho S}{4\pi^2m_pk_B}\right)^{1/3}\hbar c,
\end{equation}
where $r$ is the radius of the wind and $m_p$ is the proton mass.

\begin{figure}
    \centering
    \includegraphics[width=0.9\columnwidth]{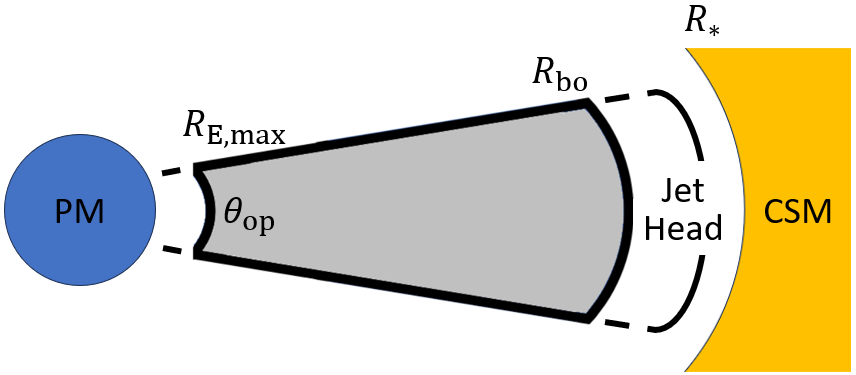}
    \caption{Simple schematic of the jet system. The protomagnetar (PM) drives the jet. We focus on regions of the jet between the jet-breakout radius, $R_{\rm bo}$, and the radius below which maximum acceleration cannot occur, $R_{\rm E,max}$. The region between these two, shaded in gray, avoids additional nuclear interactions (with the progenitor material during jet propagation) and provides a conservative region from where decay $\gamma$-ray emission may occur and escape. The jet head is located above $R_{\rm bo}$ and defines the jet region that interacts with the progenitor until the jet reaches the stellar radius, $R_*$. Above $R_*$ lies the circumstellar material (CSM).}
    \label{Jet Schematic}
\end{figure}

This neutrino-driven wind becomes magnetized, allowing processes to occur that then collimate into a jet. This collimation can occur due to oblique shocks formed when the surrounding cocoon exerts pressure on the jet \cite{Bhattacharya:2022btx}. As mass flows along open field lines, it is then able to burrow through the stellar material \cite{Metzger2011a,Bhattacharya:2022btx}. The magnetic energy stored in the wind then can be converted into bulk kinetic energy via e.g., magnetic reconnection, allowing acceleration to relativistic velocities. The jet, once it is launched and breaks out of the progenitor, will run into the CSM and begin picking up external mass which can lead to an afterglow that can outshine the nuclear decay $\gamma$-ray signal. Thus, we consider a simple description of the radius of the progenitor and its circumstellar material (CSM). This is explored in Sec.~\ref{Afterglow}.

Finally, these jets may dissipate magnetic energy that can give rise to a host of nonthermal effects. These may include the production of nonthermal photons, nuclei acceleration, and GRBs. The consequences of nonthermal effects are discussed in Sec.~\ref{Non-thermal Signal}.

\subsection{Jet properties}\label{Jet properties}

\renewcommand{\arraystretch}{1.25}
\begin{table*}[t]
    \centering
    \caption{Table for jet properties for different $B_{\rm dip}$ and $P_i$ PMs considered. The model names refer to the magnetic field strength and initial spin period. For example, ``B514P15'' refers to the model with $B_{\rm dip}=5\times10^{14}$\:G and $P_i=1.5$\:ms while ``B116P35'' refers to the model with $B_{\rm dip}=1\times10^{16}$\:G and $P_i=3.5$\:ms. Throughout this work, we assume an electron fraction of $Y_e=0.45$ for all models (however, see Sec.~\ref{Discussion} for impacts of varying $Y_e$).}
    \begin{ruledtabular}
    \begin{tabular}{ccccccc}
         Name & $B_{\rm dip}\times10^{15}$ [G] & $P_i$ [ms] & $M_\mathrm{ej}\times10^{-4}$ [M$_\odot$] & $t_\mathrm{bo}$ [s] & $t_\mathrm{E,max}$ [s] & $E_\mathrm{max}\times10^{15}$ [MeV]\footnote{$E_{\rm max}$ for an $^{56}$Fe nucleus in the jet.}\\
         \hline
         B514P15 & 0.5 & 1.5 & 21 & 3.8 & 78.6 & 1.0 \\
         B514P35 & 0.5 & 3.5 & 0.14 & 8.9 & 50.4 & 0.58 \\
         B116P15 & 10 & 1.5 & 12 & 1.4 & 38.5 & 4.0 \\
         B116P35 & 10 & 3.5 & 0.16 & 3.3 & 18.9 & 2.2 \\
    \end{tabular}
    \end{ruledtabular}
    \label{BP Models Table}
\end{table*}
\renewcommand{\arraystretch}{1}

In this work, we cover a range of dipole magnetic field strengths from $5\times10^{14}\:{\rm G}<B_{\rm dip}<1\times10^{16}\:{\rm G}$ and spins from $1.5\:{\rm ms}<P_i<3.5\:{\rm ms}$. For this range of $B_{\rm dip}$ and $P_i$, we consider the abundance of nuclei synthesized following Ref.~\cite{Ekanger:2022tia}. Since magnetized outflows may be somewhat neutron rich (see, e.g., Ref.~\cite{Reichert:2024vyd}), we consider an electron fraction of $Y_e=n_p/(n_n+n_p)=0.45$, where $n_p$ is the number density of protons and $n_n$ is the number density of neutrons. We choose this value since outflows with $Y_e\geq0.5$ will not synthesize much elements heavier than iron, as the unstable, heavy elements that give rise to $\gamma$-ray emission in large abundances are of interest. For the remainder of this paper, we take the following naming convention for our models: ``B\textit{ABC}P\textit{YZ},'' where ``\textit{ABC}'' represents $B_{\rm dip}=A\times10^{BC}$\:G and ``\textit{YZ}'' represents $P_i=Y.Z$\:ms. For example, B514P15 represents the model with $B_{\rm dip}=5\times10^{14}$\:G and $P_i=1.5$\:ms.

The mass being ejected from the PM may continue for some time. However, we consider only a subset of the jet, namely in between two epochs or radii. On the upper end, we consider the region below the breakout radius, $R_{\rm bo}$, which is the radius where the jet breaks out of the progenitor. This is because the nuclei in the jet in front of this breakout radius are subject to interactions with the stellar matter as the jet burrows through the progenitor, making nuclei survival challenging. On the other hand, nuclei in the jet below the breakout radius can escape without interacting with the stellar material on its way out. On the lower end, we consider the region above the radius at which particle acceleration can occur to $\sim 10^{15}$\:MeV (for a Fe nuclei), $R_{\rm E,max}$. Below this radius, nuclei are no longer able to be accelerated to ultrahigh energies. We adopt this since in Sec.~\ref{Non-thermal Signal} we consider the decays of accelerated nuclei. The two radii or epochs we use, as well as the mass ejected $M_{\rm ej}$ within them, are dependent on $B_{\rm dip}$ and $P_i$, whose definitions can be found in Ref.~\cite{Ekanger:2022tia}. Figure \ref{Jet Schematic} shows a schematic of the region of interest for this work, and numerical values are displayed in Table \ref{BP Models Table} as times rather than radii. Since we consider only a subset of the jet, we assume that the jet remains at a constant velocity with a bulk Lorentz factor of $\Gamma_{\rm bulk}=10$. The mass ejected during this time ($M_{\rm ej}$) is calculated by integrating the mass loss rate from $t_{\rm bo}$ to $t_{\rm E,max}$.

\subsection{Jet optical depth}\label{Opacity}

We now work out the prospects of photons escaping the jet. We quantify this by the optical depth $\tau=\int\rho_\mathrm{ej}\kappa(E_\gamma)d\ell$, where $\rho_\mathrm{ej}$ is the ejecta density, $\kappa(E_\gamma)$ is the opacity as a function of photon energy, and $d\ell$ is the path length traveled. In order to simplify the task, we assume that the jet has uniform density and that the jet is in the shape of a partial spherical sector in between $R_\mathrm{bo}(t) = \beta(t-t_\mathrm{bo})$ and $R_\mathrm{E,max}(t) = \beta(t-t_\mathrm{E,max})$, where $\beta$ is the velocity of the jet in units of the speed of light, i.e., 
\begin{equation}\label{Ejecta density}
    \rho_\mathrm{ej} = M_\mathrm{ej}\left[ \frac{2\pi}{3} \left(R_\mathrm{bo}^3 - R_\mathrm{E,max}^3\right) \left(1-\mathrm{cos}\left(\frac{\theta_{\rm op}}{2}\right)\right) \right]^{-1},
\end{equation}

\noindent where $\theta_{\rm op}\sim1/\Gamma_{\rm bulk}$ is the opening angle of the jet.

Although $\kappa(E_\gamma)$ depends on both the isotopic composition of the jet and the photon energy, we assume $\kappa$ takes a constant value for simplicity. We choose a value of 0.1\:cm$^{2}$\:g$^{-1}$, from Fig.~6 of Ref.~\cite{Chen:2021tob}, which is motivated by the slightly neutron-rich ``blue kilonovae'' in Ref.~\cite{Waxman:2017sqv}. Note that the model only extends up to $Y_\mathrm{e}=0.4$, whereas our jets assume $Y_\mathrm{e}=0.45$. We thus use the value for $Y_\mathrm{e}=0.4$, but the opacity shows only a slow variation and we do not believe that this would greatly increase. With this setup, the optical depth takes the form 
\begin{equation}\label{Optical depth}
    \tau = \rho_\mathrm{ej} \: \kappa \: (R_\mathrm{bo} - R_\mathrm{E,max}).
\end{equation}

\begin{figure}[t]
    \centering
    \includegraphics[width=\linewidth]{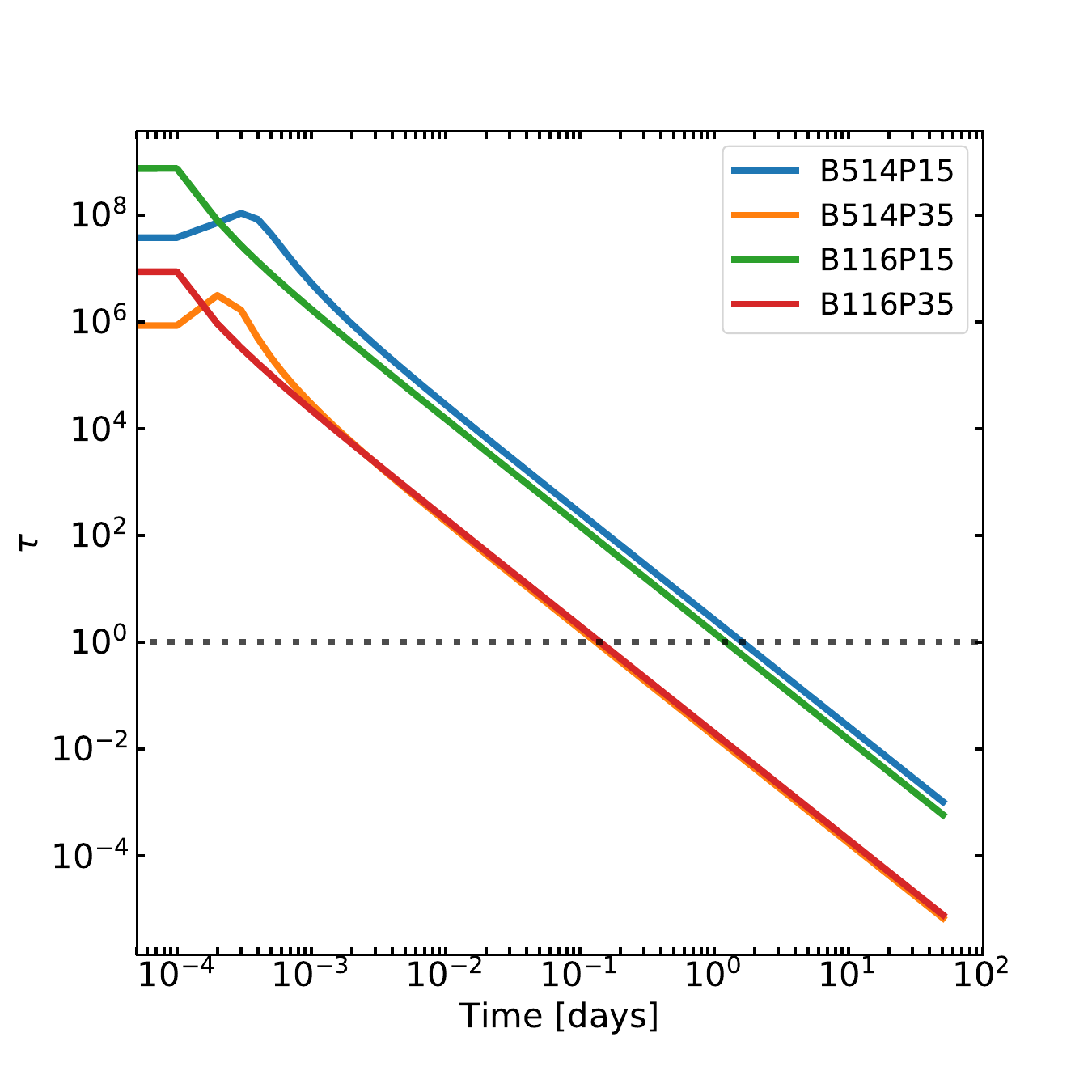}
    \caption{Optical depth ($\tau$) for escaping gamma rays as a function of time for the different PM models considered in this work. The primary source of this opacity is from the ejecta itself. The largest factor on how this opacity evolves is the amount of mass ejected in each model, so the highly spinning models with $P_i=1.5\:{\rm ms}$ are optically thick for the longest: up to $\sim2\:{\rm days}$.}
    \label{Optical Depth Plot}
\end{figure}

We carry out the optical depth calculation for the four different models shown in Table \ref{BP Models Table} and plot the different optical depths as a function of time in Fig.~\ref{Optical Depth Plot}. Note that we only consider the opacity due to the rest of the jet ejecta. 

We see that the largest difference comes from models with different $P_i$ values, which in turn comes from the significantly different $M_\mathrm{ej}$. The models with smaller periods become optically thin after 2--3\:days, whereas the larger period models become optically thin after 0.1--0.2\:days. The authors of Ref.~\cite{Chen:2021tob} calculate the opacity at $t=1$\:day after nucleosynthesis, which is roughly the timescale at which the jets become optically thin, so the assumption that $\kappa=0.1$\:cm$^{2}$\:g$^{-1}$ seems reasonable. 

\subsection{Afterglow timing}\label{Afterglow}

In order to determine the impacts of the afterglow, we determine when the jet should enter the afterglow phase. The timing of the afterglow phase is represented by the condition that the jet has swept through a mass of material equal to the mass of the jet. Therefore, to find this we must adopt some model for the CSM around the progenitor. 

We consider two cases for the CSM, one in which there is a flat density profile (case 1) and another in which there is a dense ejecta shell formed by pre-SN mass-loss with the profile $\rho_\mathrm{CSM}(r) = 5\times10^{16}\: \mathrm{g} \: \mathrm{cm}^{-3} \: D_* \: r^{-2}$ from the stellar radius, $R_*$, out to $10^{16}$\:cm \cite{Murase:2017CSM}, followed by a flat density profile (case 2). In both cases, we assume that the constant density portion of the CSM has a mass density of $\rho_\mathrm{CSM}=10^{-25}$\:g\:cm$^{-3}$ (roughly 0.1 protons per cm$^3$). This is motivated by simulations of the CSM around massive stars, e.g., Refs.~\cite{Georgy:2013CSM, Dwarkadas:2022CSM, Dwarkadas:2023CSM}. As this is highly dependent on the progenitor, we choose a somewhat conservative (i.e., large, which brings about an earlier afterglow phase) value for the constant density. For the second case, we adopt two models of pre-SN mass-loss from Ref.~\cite{Murase:2017CSM}, specifically the model with maximal pre-SN mass-loss (IIn) with $D_*=1$ and $R_*=10^{13}$\:cm and the model with minimal pre-SN mass-loss (Ibc) with $D_*=10^{-5}$ and $R_*=3\times10^{11}$\:cm.

With the density profiles of the CSM defined, we can now calculate the mass picked up by the jet as it propagates through the CSM. For this, we assume that the jet moves with a constant opening angle and picks up all of the CSM that it encounters. The mass of CSM swept through is then just found by integrating the density profile over the volume the jet has traveled, i.e., 
\begin{equation}\label{CSM mass encounterd}
    M_\mathrm{enc}(t) = 2\pi \left[1-\mathrm{cos}\left(\frac{\theta_{\rm op}}{2}\right) \right]\int^{r(t)}_{R_*} dr' \: r'^2 \rho_\mathrm{CSM}(r'),
\end{equation}
\noindent where the prefactor comes from carrying out the angular portions of the volume integral, $r(t)$ is how far the jet has traveled at time $t$, $R_*$ is the stellar radius, and $\rho_\mathrm{CSM}(r)$ is the CSM density profile. We can then solve for $t$ to find the time that the afterglow should occur by setting $M_\mathrm{enc}(t)=M_\mathrm{ej}$. Using the values of $M_\mathrm{ej}$ from Table \ref{BP Models Table}, we find the afterglow timing for cases 1 and 2, which are shown in Table \ref{Afterglow Table}.

We see that there are vast differences between case 1 and the IIn model of case 2 for when the afterglow phase begins. However, the afterglow timing for Case 1 and the Ibc model of Case 2 are the same. This is a result of the large amount of pre-SN mass loss for the IIn progenitors, while the Ibc progenitors have little pre-SN mass loss. In the IIn pre-SN mass-loss model, the jet would sweep up $\sim6\times10^{-3}\:\mathrm{M}_\odot$ before leaving the pre-SN mass-loss based CSM shell. This is greater than any of the $M_\mathrm{ej}$ of the jet models we consider, therefore the jet should enter the afterglow phase within the pre-SN mass loss CSM shell. In the Ibc mass-loss model, the jet will only sweep through $\sim10^{-7}\:\mathrm{M_\odot}$ of material within the pre-SN mass-loss shell, which is much smaller than any $M_\mathrm{ej}$ considered, therefore mass swept through is dominated by the outer constant density CSM. Note, we do not expect the afterglow phase to begin after $\mathcal{O}(10^3)$\:days, as we extend the low density CSM to infinity, which is not realistic. However, a more realistic and complex model of the CSM surrounding the progenitor star is beyond the scope of this work.
\renewcommand{\arraystretch}{1.25}
\begin{table}[t]
    \caption{Times at which each jet model enters the afterglow phase, in days. The different cases refer to different density profiles of the CSM. Case 1 assumes a flat density profile and case 2 assumes a dense shell of pre-SN mass out to some radius with a flat density profile outside of the shell.}
    \begin{ruledtabular}
    \centering
    \begin{tabular}{cccc}
         \multicolumn{1}{c}{Model} & \multicolumn{1}{c}{Case 1} & Case 2: IIn & Case 2: Ibc \\
         & [days] & [days] & [days] \\ \hline
         B514P15 & $6.75\times10^3$ & $1.37$ & $6.75\times10^3$ \\
         B514P35 & $1.27\times10^3$ & $0.0130$ & $1.27\times10^3$ \\
         B116P15 & $5.60\times10^3$ & $0.782$ & $5.60\times10^3$ \\
         B116P35 & $1.33\times10^3$ & $0.0143$ & $1.33\times10^3$ \\
    \end{tabular}
    \label{Afterglow Table}
    \end{ruledtabular}
\end{table}
\renewcommand{\arraystretch}{1}

\section{Population of unstable nuclei}\label{Heavy Nuclei}

We discuss the nuclei contents of PM jets. For this, we first compute nucleosynthesis yields, then discuss nuclei survival as the wind evolves. 

\subsection{\ensuremath{r}-process nucleosynthesis}\label{Nucleosynthesis}

To obtain the detailed distributions of the abundance of nuclei synthesized in winds, we use the results of Ref.~\cite{Ekanger:2022tia}, which models the density and temperature evolution of PM winds over the same $B_{\rm dip}$, $P_i$, and $Y_e$ choices as in this work. These density and temperature evolution curves are then input to the nuclear reaction network {\tt SkyNet} \cite{Lippuner:2017tyn} to calculate nuclear abundances as a function of time. We assume that the system is initially in nuclear statistical equilibrium (NSE), composed primarily of neutrons and protons and consistent with a $Y_e=0.45.$ We then use the forward reaction rates from the REACLIB database \cite{reaclib} and use detailed balance to calculate inverse rates to be consistent with NSE. This large network tracks 7836 species up to mass numbers of $A=337$ and the network is evolved until $\sim100$\:s postbounce, although the nucleosynthesis occurs over a timescale of $\sim10$\:ms (e.g., Fig. 2 of Ref.~\cite{Ekanger:2022tia}).

\begin{figure}[t]
    \centering
    \includegraphics[width=\linewidth]{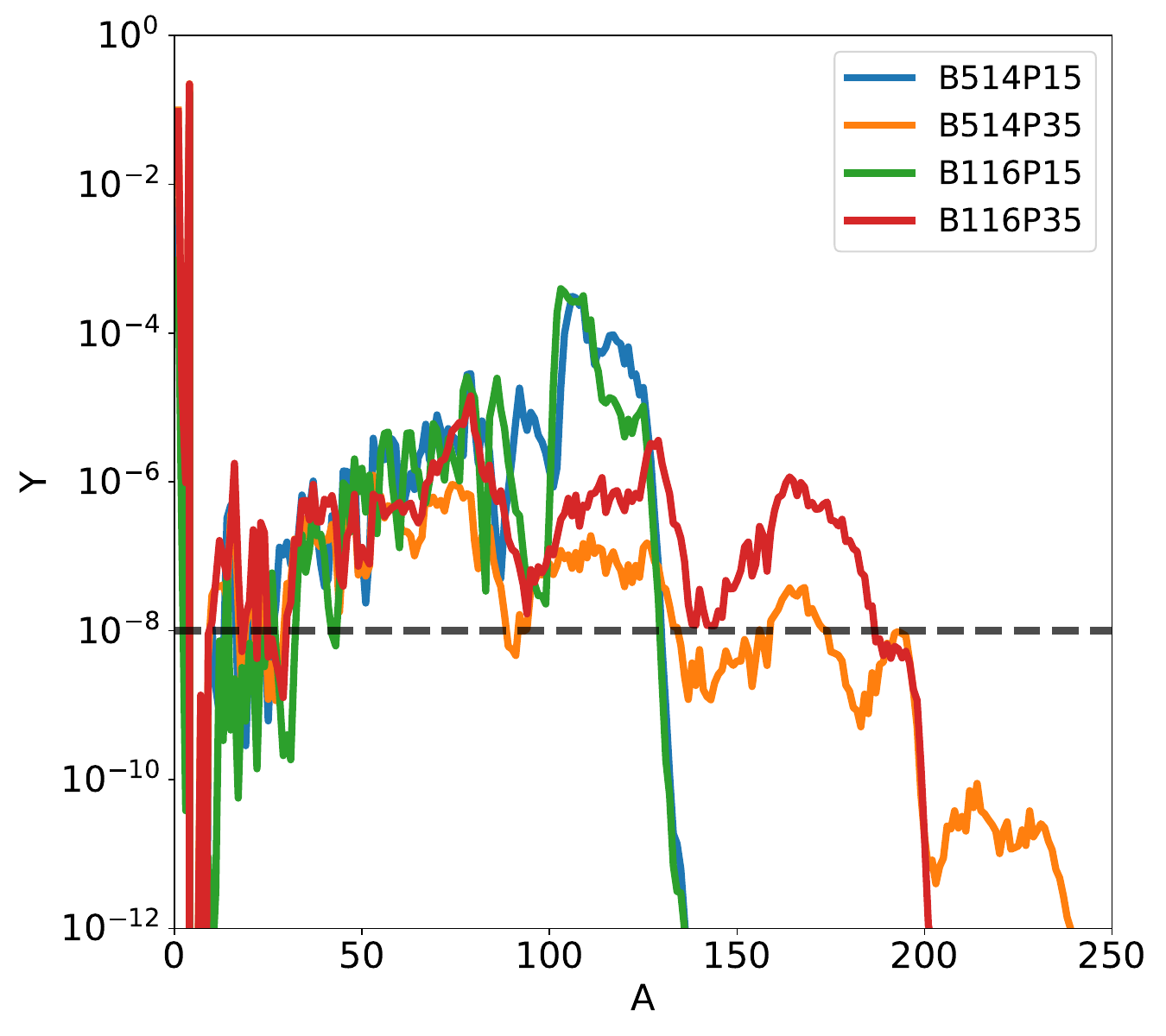}
    \caption{Abundances for the PM models used in this work. The black dashed line at $Y=10^{-8}$ represents our chosen cutoff for computing decay $\gamma$-ray signals. Note that the $P_i=1.5\:{\rm ms}$ models produce elements peaked around the first $r$-process peak and are larger in abundance than the $P_i=3.5\:{\rm ms}$ models. Although the latter produces nuclei with higher mass numbers, they are produced in lower abundance compared to the first peak nuclei.}
    \label{Model Abundances}
\end{figure}

Figure \ref{Model Abundances} shows the abundances ($Y$) of nuclei synthesized as a function of their mass number ($A$). This abundance distribution is shown for a representative time (labeled $t_{\rm photo}$, based on Ref.~\cite{Ekanger:2022tia}), which is between $t_{\rm bo}$ and $t_{\rm E,max}$. Although each model has the same $Y_e=0.45$, the different $B_{\rm dip}$ and $P_i$ play a role in the distribution of the abundances. Each model can undergo the ``weak'' $r$ process, i.e., synthesize nuclei above the first $r$-process peak. However, the lesser $P_i$ models synthesize nuclei with $\sim100<A<130$ in greater abundance, while the greater $P_i$ models synthesize heavier mass numbers but in lesser abundance. Models with shorter $P_i$ values increase the mass loss rate due to centrifugal slinging, but also suppress the entropy. These increase the nucleon density and (somewhat) decrease the temperature, respectively, resulting in the abundance pattern shown. Many of these heavy nuclei are unstable to decays on timescales much longer than the nucleosynthesis timescale and are, therefore, prime targets for $\gamma$-ray searches. Also shown in Fig.~\ref{Model Abundances} is a dashed black line denoting the abundance cutoff of $Y=10^{-8}$, below which we assume the abundance is too low to produce a detectable $\gamma$-ray signal.

\subsection{Survival of heavy nuclei}

Heavy nuclei are synthesized when the wind falls out of NSE, but may be exposed to additional destructive processes on their way out of the star. For example, if photodisintegration breaks nuclei into protons and neutrons on a timescale shorter than the decay half-lives of unstable nuclei, this could eliminate any potentially detectable decay $\gamma$-ray signal. Therefore, we must assess whether or not nuclei can survive the wind environment on their way out of the star.

Nuclei survival depends on the nature of the ambient photon field, the wind model, and the energy of the nuclei. The nature of the photon field depends on the evolution of the Thomson optical depth; if $\tau_T\gg1$ electrons are thermalized to the same temperature as the photons whereas photons are nonthermal if $\tau_T\ll1$. We assume many modeling aspects of Ref.~\cite{Ekanger:2022tia}, where a nonthermal photon spectrum is assumed. Analytical estimates from Ref.~\cite{Carpio:2023wyt} suggest, though, that the Thomson optical depth below the jet head is $\gg1$.

We treat nuclei photodisintegration and survival in the following way. Thermal nuclei are typically not photodisintegrated regardless of whether the ambient photon field is thermal or nonthermal [Ekanger \textit{et al.} 2025 (to be published)]. If nuclei are accelerated to very high energies however, photodisintegration will occur against the thermal photon field. We focus on the case where the jet is relativistic with Lorentz factor $\Gamma_\mathrm{bulk}=10$, but the nuclei and photons are thermalized in the jet comoving frame. This motivates that heavy nuclei survive disintegration in the jet frame. In Sec.~\ref{Non-thermal Signal} we consider the scenario where some particle acceleration occurs. In this case, we must assume nuclei survival---which can occur if photons are also nonthermal---but we do not investigate the process in detail in this work.

\section{Thermal Nuclei}\label{Signal}

We generate line signals from the unstable thermal nuclei within the PM jet. The $\gamma$ rays originate from $\beta^\pm$ decays inside the jet. Under the assumption that the jet is pointed towards us, the signal is enhanced via relativistic beaming. We produce light curves as well as unsmoothed energy spectra visible at Earth taking into account the optical depth within the jet. 

\subsection{Decays}\label{Decays}

In order to model the $\gamma$-ray signal from the decay of unstable nuclei boosted in the jet, we only model decays that produce $\gamma$ rays, e.g., $\beta^\pm$. We also only track the $\gamma$ rays produced from the decays of the first generation of unstable nuclei and their daughter nuclei. In order to find the rate of photon production, we need the radioactive decay equations for a parent and daughter nucleus, but only including the terms that result in the respective $\gamma$ emission for each, i.e., 
\begin{equation}\label{Photon Rates}
    \begin{split}
        \frac{dN_\mathrm{p,\gamma}}{dt} &= \lambda'_\mathrm{p} I_{\gamma,\mathrm{p}} N_\mathrm{p}(t_\mathrm{E,max}) e^{-\lambda'_\mathrm{p}t}, \\
        \frac{dN_\mathrm{d,\gamma}}{dt} &= \frac{\lambda'_\mathrm{p}\lambda'_\mathrm{d}}{\lambda'_\mathrm{d}-\lambda'_\mathrm{p}} I_{\gamma, \mathrm{d}} N_\mathrm{p}(t_\mathrm{E,max}) (e^{-\lambda'_\mathrm{p}t} - e^{-\lambda'_\mathrm{d}t}),
    \end{split}
\end{equation}
\noindent where $t$ is the time of measurement after $t_\mathrm{E,max}$, $\lambda'_i$ are the time-dilated decay constants ($\lambda'_i\equiv\Gamma_\mathrm{bulk}\lambda_i, \: \lambda_i\equiv\mathrm{ln}(2)/\tau_{1/2, i}$ where $\tau_{1/2, i}$ is the half-life), $I_{\gamma, i}$ are the $\gamma$-ray intensities, and $N_\mathrm{p}(t_\mathrm{E,max})$ is the initial number of parent nuclei. The nuclear data were taken from the decay radiation search page of the NUDAT database\footnote{\url{https://www.nndc.bnl.gov/nudat3/indx_dec.jsp}}, version 3 \cite{Nudat}. $N_\mathrm{p}(t_\mathrm{E,max})$ is found directly from the abundances and the total ejecta mass, as described in Sec.~\ref{Heavy Nuclei}.

Note that we exclude $\gamma$ rays from metastable states through isomeric transitions. Because the nuclear network we use does not track how the nuclei spin states are populated and the NUDAT database does not provide this information, we do not estimate how these $\gamma$-ray lines would evolve within our system. These isomeric transitions, however, could be important signals and we discuss their potential impact in Sec.~\ref{Discussion}.

\subsection{Measured signal}\label{Measured Signal}

\begin{figure}[t]
    \centering
    \includegraphics[width=\linewidth]{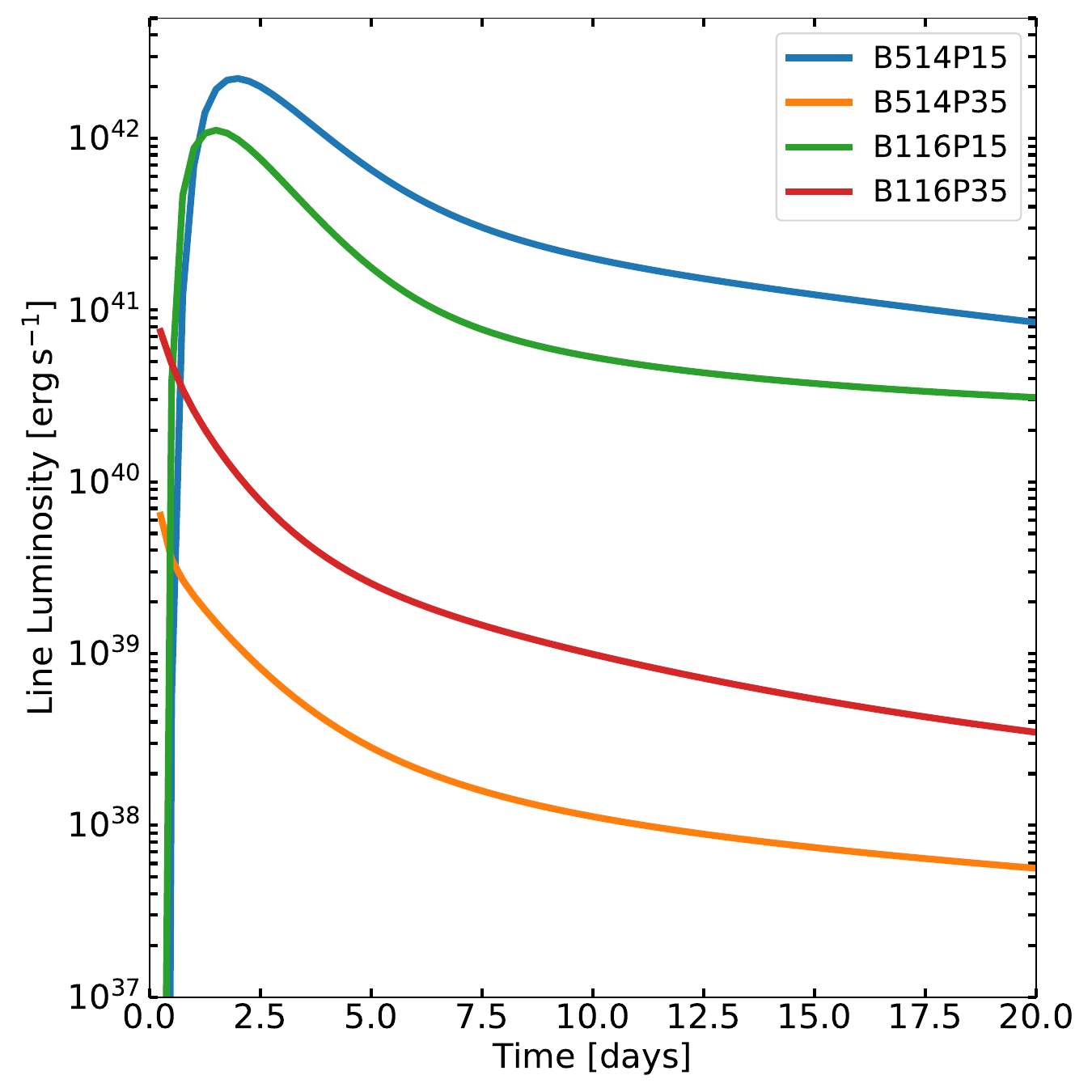}
    \caption{Light curves of decay lines for the four different models considered in this work. The luminosity is computed by summing the total instantaneous energy emission rate from $\gamma$-ray lines. The luminosity peaks when the jet becomes optically thin, $t<1$\:day for $P_i=3.5$\:ms models and $t>1$\:days for $P_i=1.5$\:ms models. This assumes the jet pointed towards Earth ($\theta_\mathrm{view}=0^{\circ}$) such that relativistic beaming maximally boosts the luminosity.}
    \label{Total Light Curve}
\end{figure}

\begin{figure*}[t]
    \centering
    \includegraphics[width=0.45\linewidth]{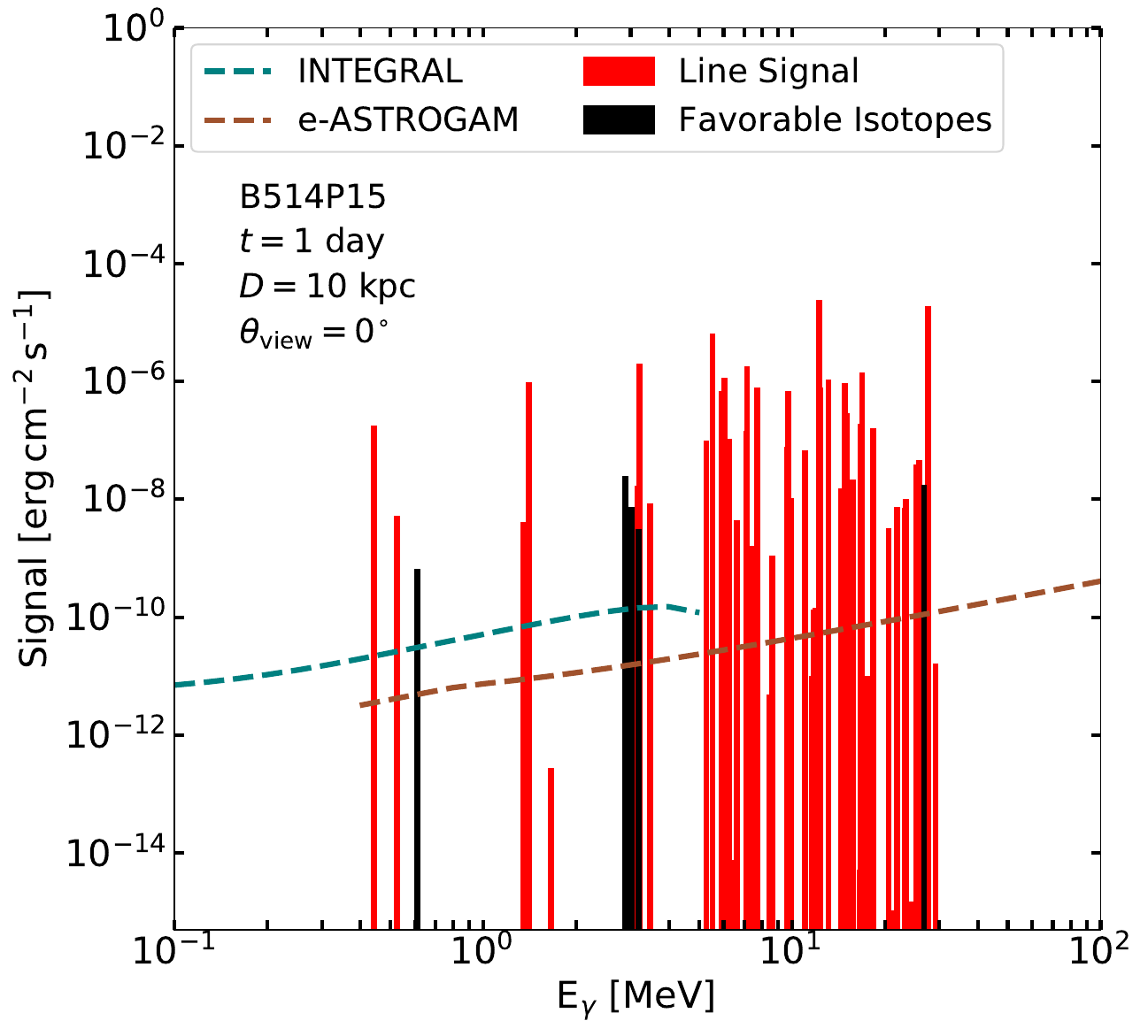}
    \includegraphics[width=0.45\linewidth]{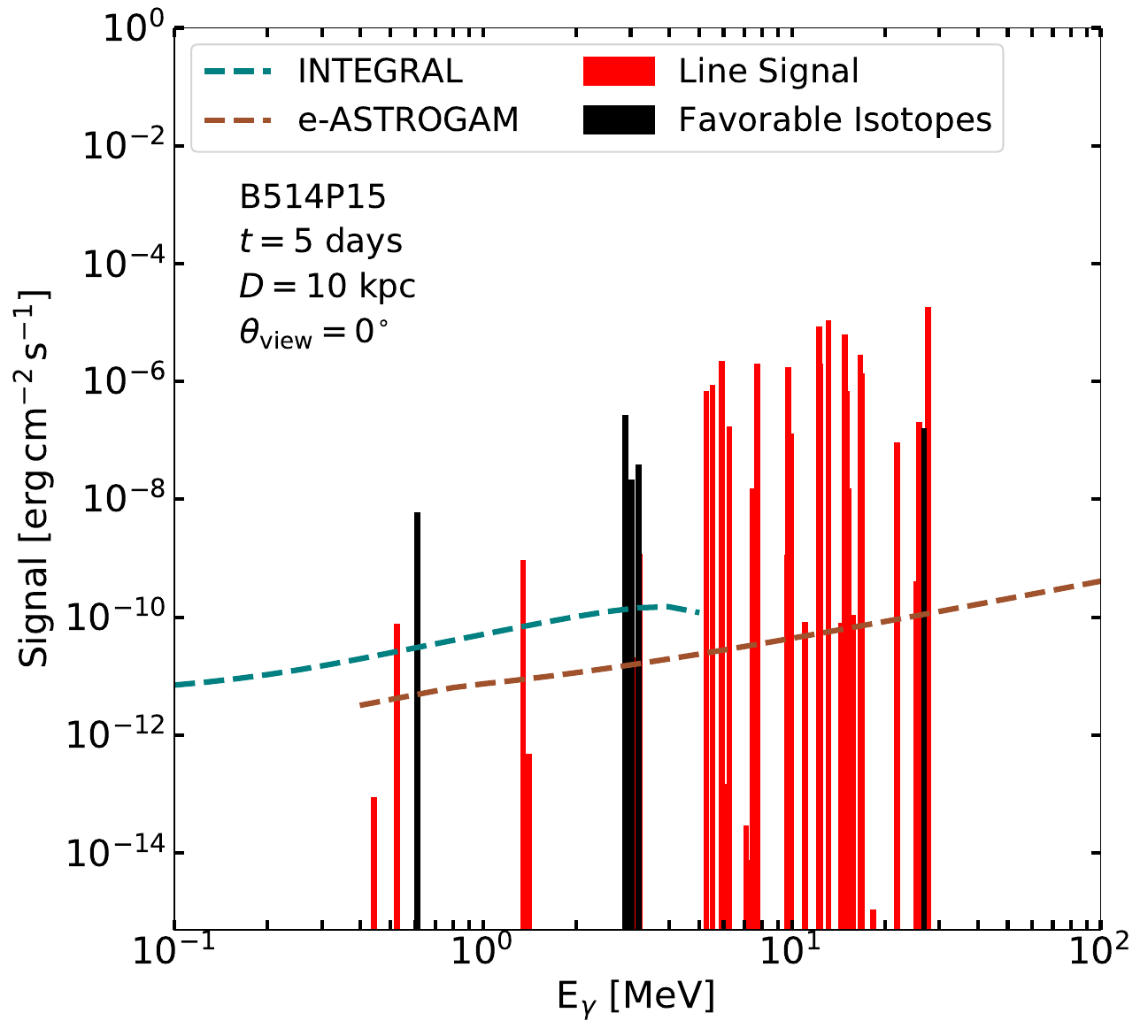}
    \\
    \includegraphics[width=0.45\linewidth]{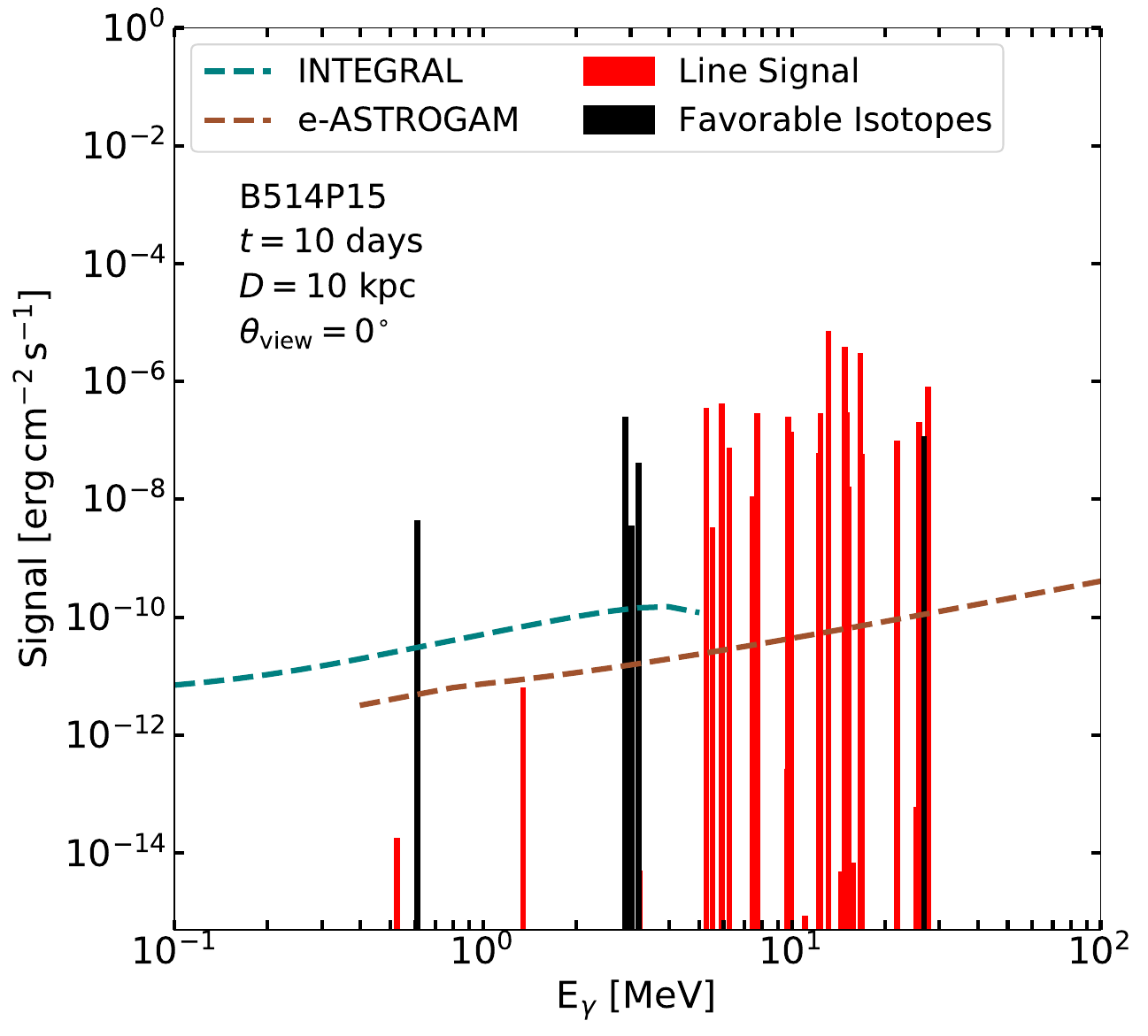}
    \includegraphics[width=0.45\linewidth]{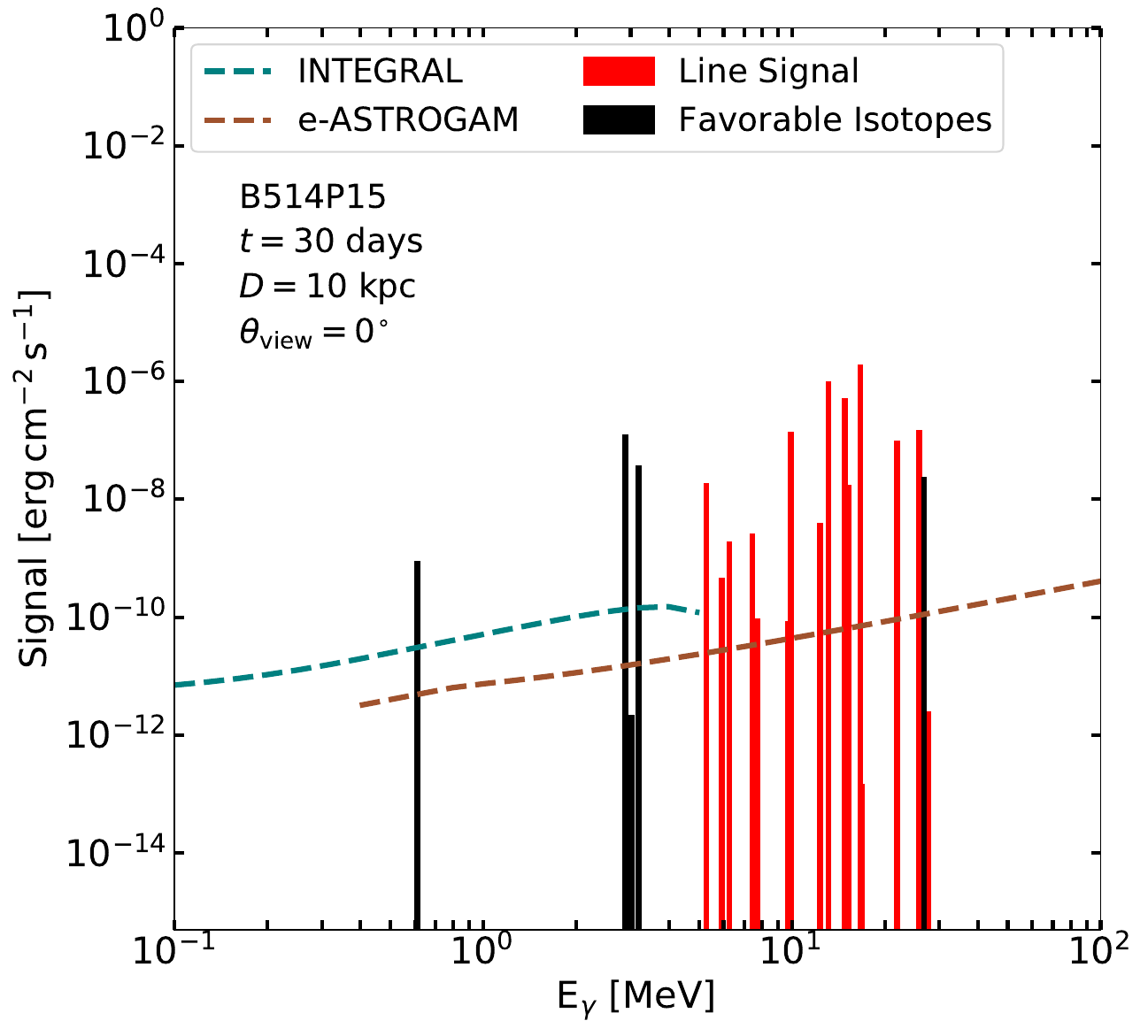} 
    \caption{Line emission from the decays of thermal nuclei, for the case of $B_\mathrm{dip}=5\times10^{14}$ G, $P_i=1.5$ ms, and $\Gamma_\mathrm{bulk}=10$. The four panels show the spectrum at different times of evolution: $t=1$ day (top left), $t=5$ days (top right), $t=10$ days (bottom left), and $t=30$ days (bottom right). Dashed lines show detector point-source line sensitivities for an observation time of $10^6$\:s. The plot uses the energy resolution of \textit{INTEGRAL/SPI}, for which we adopt $E/\Delta E \sim 450$, therefore there is little blending due to the fineness of the energy resolution. In red is the line signals for the parent and daughter isotopes that meet the abundance cutoff criterion (see dashed line in Fig.~\ref{Model Abundances}). In black are a sub sample highlighting long-lived isotopes that are visible in both \textit{INTEGRAL/SPI} and \textit{e-ASTROGAM} for ten days (see Table \ref{Important Isotopes}). Note here also that there is a black line at $1.3$\:MeV, which is another line from the $^{28}$Mg isotope which is among the favorable list of isotopes (see Table~\ref{Important Isotopes}). }
    \label{Thermal Line Emission Over Time}
\end{figure*}

\begin{figure*}[t]
    \centering
    \includegraphics[width=0.45\linewidth]{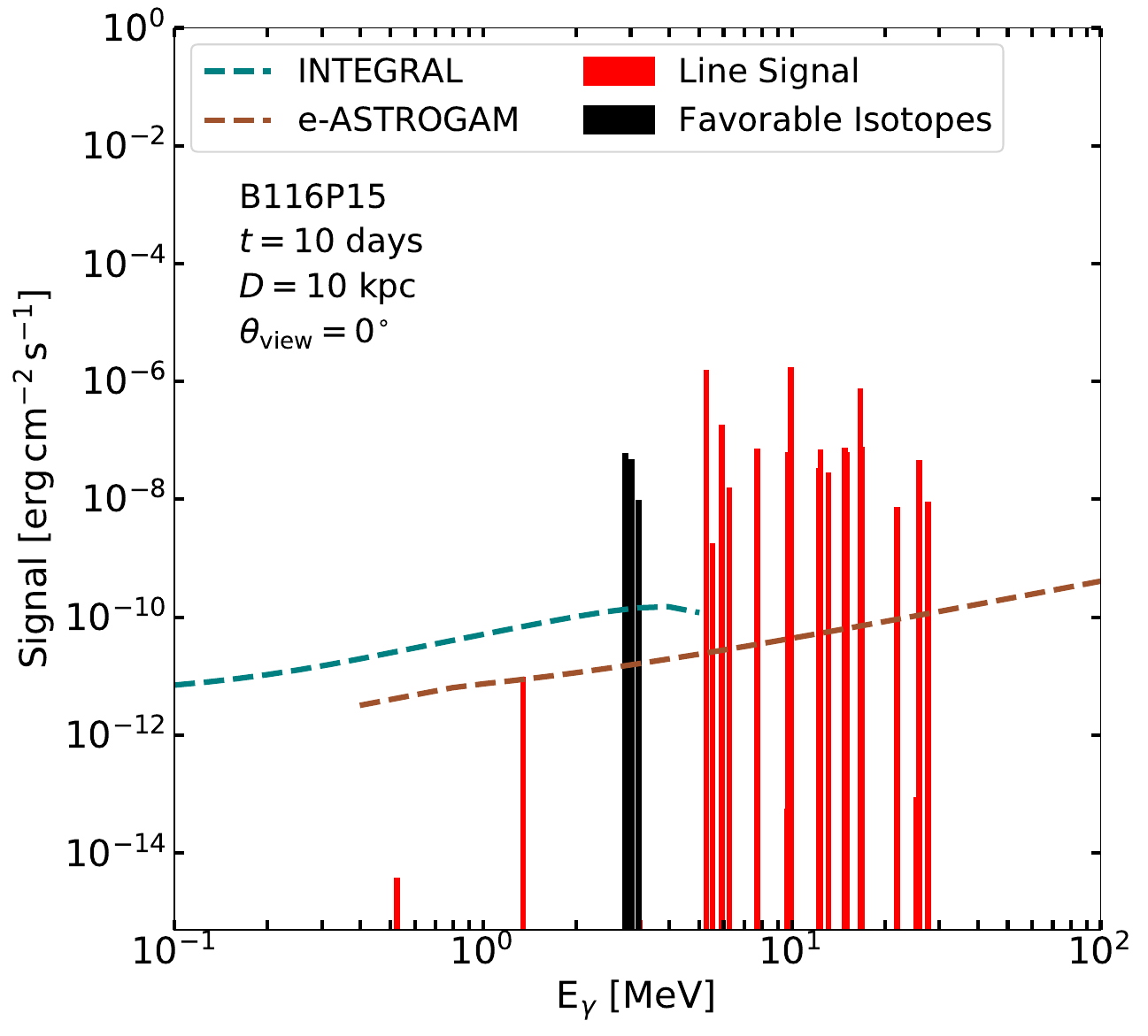}
    \includegraphics[width=0.45\linewidth]{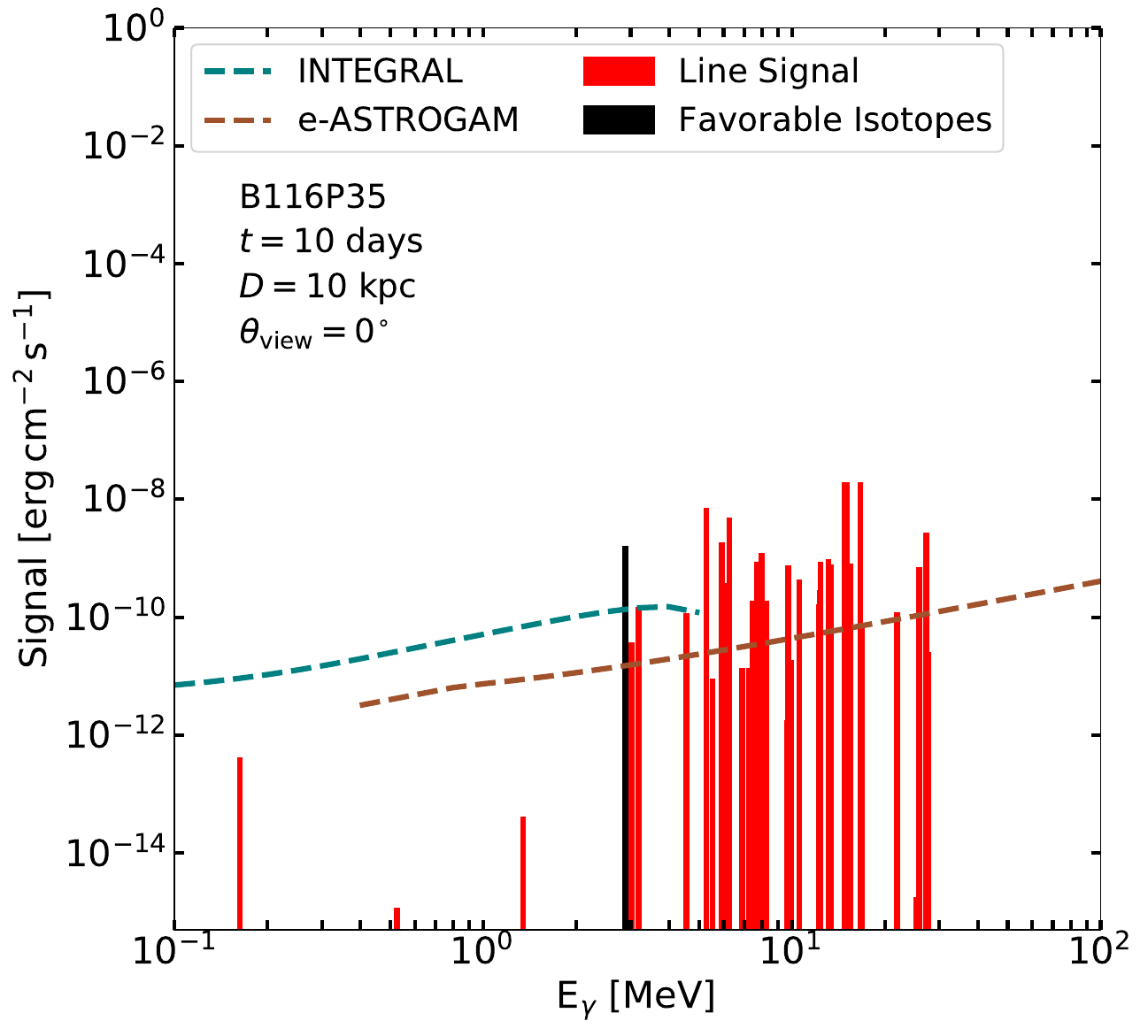}
    \\
    \includegraphics[width=0.45\linewidth]{Lines_B514P15_Day10.pdf}
    \includegraphics[width=0.45\linewidth]{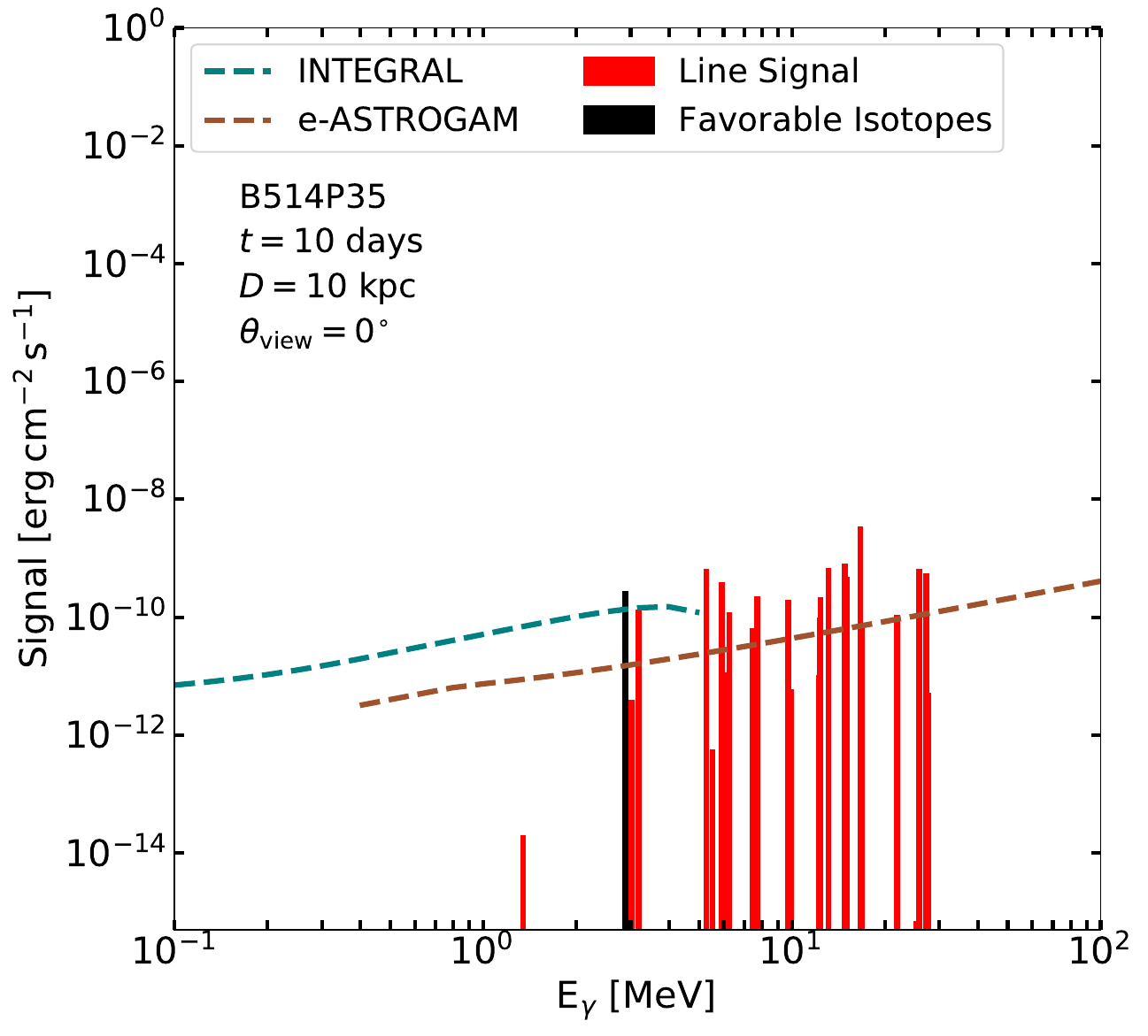}
    \caption{Same as Fig.~\ref{Thermal Line Emission Over Time} but showing the different $B_\mathrm{dip}$ and $P_i$ models used in this work, all at $t=10$ days at which point all models are optically thin and allow $\gamma$-ray lines to escape. The columns correspond to matching $P_i$ and the rows correspond to matching $B_\mathrm{dip}$. There is a stronger dependence on the period than the magnetic field strength for the relative line signals. Again, dashed lines show detector point-source line sensitivities for an observation time of $10^6$\:s}
    \label{Differing Line Emission t=10 days}
\end{figure*}

With the $\gamma$-ray production rates from in Eq.~(\ref{Photon Rates}), we can calculate the predicted $\gamma$-ray signal for the different $B_\mathrm{dip}$ and $P_i$ models we consider in this work. First, we need to fold in the fact that the photons have to escape the jet, which is described by the optical depth found in Sec.~\ref{Opacity}. This affects the observed $\gamma$-ray rates in the following way: 
\begin{equation}
    \frac{dN_{i,\gamma}^\mathrm{observed}}{dt} = \frac{dN_{i,\gamma}}{dt}\: e^{-\tau}.
\end{equation}
\noindent As the optical depths, and therefore densities, are small on timescales longer than a day, we do not consider down-scattering of the photons due to reprocessing and interactions within the jet. These would be important for epochs where $\tau \gg 1$, which is not the focus of this work. 

Using these observed photon rates and the specific energy of each $\gamma$ from the individual nuclear decays, we can predict the $\gamma$-ray light curves for each model. To do this, we sum the luminosity of the $\gamma$ rays produced from parent and daughter decays at each time step, i.e., 
\begin{equation}
    L(t) = \delta^3\sum_{\mathrm{isotopes}} \frac{dN_{i,\gamma}^\mathrm{observed}}{dt} E_{i,\gamma}, 
\end{equation}
\noindent where $\delta$ is the Doppler factor [$\delta=[\Gamma(1-\beta\:\mathrm{cos}(\theta_\mathrm{view}))]^{-1}$ for $\theta_\mathrm{view}$ being the angle between the jet propagation direction and the line of sight] and $E_{i, \gamma}$ is the specific $\gamma$-ray line energy for the decay of the $i$th unstable isotope. The $\delta^3$ term is the result of relativistic beaming; see Appendix B of Ref.~\cite{Urry_1995} for a derivation. We plot these light curves for the different cases of $B_\mathrm{dip}$ and $P_i$ in Fig.~\ref{Total Light Curve}. We see that the light curves are quite different for the different cases, where the larger $P_i$ cases have much lower luminosities, by 2--3 orders of magnitude. This comes from the fact that there is less mass ejected for these higher periods and that there are less unstable nuclei formed during nucleosynthesis. The magnetic field strength also changes the luminosity as it also impacts the mass ejected, but as a second-order effect that can still be quite strong ($\gtrsim 1$ order of magnitude).

The time at which the $\gamma$-ray light curve reaches its peak luminosity is determined by the time at which the jet becomes optically thin, which is shown in Fig.~\ref{Optical Depth Plot}. In turn, the time at which the jet becomes optically thin is directly proportional to the mass of the jet, which itself depends mainly on the value of $P_i$. 

To explore if these line signals are observable by $\gamma$-ray detectors that are currently working or are planned for the near future, we compute the energy flux, which is both a function of time and the specific energy bin considered. As most nuclear decays produce $\gamma$'s in the 10\:keV to 1\:MeV energy range, we focus on detectors that are sensitive to 200\:keV to 20\:MeV $\gamma$'s as we assume $\Gamma_\mathrm{bulk}=10\Rightarrow\delta\sim20$ for when the jet is pointed towards us. Therefore, we consider \textit{INTEGRAL/SPI} \cite{Integral:2003nn} for a current detector and \textit{e-ASTROGAM} \cite{e-ASTROGAM:2016bph} for a future detector. For illustration, we bin the signal with the energy resolution of \textit{INTEGRAL/SPI} ($E/\Delta E \sim 450$)\footnote{See\;\url{https://www.cosmos.esa.int/web/integral/instruments-spi}.}, which is sensitive enough to avoid blending of individual lines. \textit{e-ASTROGAM} will have a worse energy resolution ($\sim3$\%) \cite{e-ASTROGAM:2016bph}, which could cause some line blending.

\renewcommand{\arraystretch}{1.25}
\begin{table}[t]
    \centering
    \caption{Favorable isotopes among the list of all unstable nuclei that give rise to $\gamma$-ray signals in our modeling. Here, ``Model'' refers to the central engine properties, ``Sym.'' is the chemical symbol for the favorable isotope, $\tau_{1/2}$ is the rest-frame half-life of that isotope, and $E_{\gamma}$ is the rest-frame energy of the decay line. Nuclei are categorized as favorable if they are within the sensitivity of both \textit{INTEGRAL} and \textit{e-ASTROGAM} at ten days after $t_\mathrm{E,max}$.}
    \begin{ruledtabular}
    \begin{tabular}{cccccc}
         Model & Sym. & $\tau_{1/2}$ [s] & $E_\gamma$ [MeV]\\
         \hline
         \multirow{4}*{B514P15} & $^{28}$Mg & $7.53\times10^5$ & $0.031$\\
         & $^{47}$Sc & $2.89\times10^5$ & $0.159$ \\
         & $^{72}$Zn & $1.67\times10^5$ & $0.145$ \\
         & $^{85}$Kr & $1.61\times10^4$ & $0.151$ \\
         \hline
         B514P35 & $^{47}$Sc & $2.89\times10^5$ & $0.159$ \\
         \hline
         \multirow{3}*{B116P15} & $^{47}$Sc & $2.89\times10^5$ & $0.159$ \\
         & $^{72}$Zn & $1.67\times10^5$ & $0.145$ \\
         & $^{85}$Kr & $1.61\times10^4$ & $0.151$ \\
         \hline
         B116P35 & $^{47}$Sc & $2.89\times10^5$ & $0.159$ \\
    \end{tabular}
    \end{ruledtabular}
    \label{Important Isotopes}
\end{table}
\renewcommand{\arraystretch}{1}

The energy flux that a detector would observe is then
\begin{equation}
    E_\gamma\Phi(E\pm\Delta E, t) = \frac{L(E\pm\Delta E,t)}{4\pi D^2}, 
\end{equation}
\noindent where $L(E\pm\Delta E,t)$ is the luminosity within the energy bin considered and $D$ is the distance to the source. To test the signal from a galactic event, we assume $D=10$\:kpc, which is roughly the distance to the Galactic Center, where it is most probable for a CCSN to occur. In Fig.~\ref{Thermal Line Emission Over Time}, we plot the thermal line signal at four different times ($t=$\:1, 5, 10, and 30\:days) for the model where $B_\mathrm{dip}=5\times10^{14}$\:G and $P_i=1.5$\:ms. We see that for $t\lesssim30$\:days, there are several lines that are visible with \textit{INTEGRAL}. These long-lasting lines come from $^{28}\mathrm{Mg}$, $^{47}\mathrm{S}$, $^{72}\mathrm{Zn}$, and $^{85}\mathrm{Kr}$ (see Table \ref{Important Isotopes}). Because of the boosting of the line energies, most of them are only visible to detectors like \textit{e-ASTROGAM}, that have sensitivities for $E_\gamma\gtrsim10$\:MeV. We note that there are a few lines which are visible out to $t=10$\:days in both detectors, which are shown in black and are described in Table~\ref{Important Isotopes}. 

The line signal at $t=10$\:days for the different models is plotted in Fig.~\ref{Differing Line Emission t=10 days}, where the columns correspond to the same value for $P_i$ and the rows correspond to the same value for $B_\mathrm{dip}$. The difference between the models where $P_i=1.5$\:ms is apparent when looking at the right end of the signal, where the model with $B_\mathrm{dip}=5\times10^{14}$\:G has stronger signals above 1\:MeV. The difference is much more apparent for the models where $P_i=3.5$\:ms, as the overall signal strengths are mostly detectable with \textit{e-ASTROGAM} at later times.

Because the jet is beamed and the $\gamma$-ray signal is significantly enhanced, the detection horizon for on-axis jets ($\theta_{\rm view}=0^{\circ}$) is extragalactic. We define this detection horizon as the distance at which the last decay line is visible within \textit{INTEGRAL} and \textit{e-ASTROGAM} sensitivity. Since some lines are particularly strong at $\sim10$\:MeV, the distance horizon for each model goes as follows: $\sim35$\:Mpc for B514P15, $\sim21$\:Mpc for B116P15, $\sim0.7$\:Mpc for B514P35, and $\sim3$\:Mpc for B116P35. Off-axis jets, however, will be much more difficult to see and are limited to well within the galaxy. The critical viewing angle for observation at the galactic center, i.e., the angle at which the signal disappears when placed at $10$\:kpc, goes as follows: $\theta_{\rm view}\sim70^\circ$ for B514P15, $\sim25^\circ$ for B514P35, $\sim65^\circ$ for B116P15, and $\sim33^\circ$ for B116P35. Given the fairly wide viewing angles, there is still reasonable possibility for a detectable signal. 

\section{Nonthermal Nuclei}\label{Non-thermal Signal}

Magnetized winds may dissipate energy such that nuclei are accelerated to, potentially, ultra-high energies. If nuclei can be accelerated without being disintegrated into nucleons, the half-lives and decay photon energies of unstable heavy elements may be boosted considerably. In this section, we examine how some fraction of accelerated nuclei give rise to a long-lived, high-energy $\gamma$-ray tail that could be detectable across many experiments. The following section assumes that the nuclei survive the non-thermal acceleration. However, the validity of this requires further work. The survival of accelerated high-energy nuclei has been shown to be possible for compact progenitors like the model considered in this work, but only at late times approaching $\sim100$\:s [Ekanger \textit{et al.} 2025, (to be published)]. Furthermore, the investigation focused on nuclei with energies up to $\sim10^{14}$\:eV, so the survival of even higher energy nuclei is not clear. Thus, the continuum signal predicted from this section should be considered an upper limit to the potential signal. This is also discussed in models of Refs.~\cite{Wang:2008grb,MGH11,Horiuchi:2012grb, Bhattacharya:2021cjc}. We also do not consider the overlap with the afterglow, as this is very progenitor dependent. 

\subsection{Continuum signal}\label{Continuum Signal}
If non-thermal acceleration occurs in the jets we model, there would be a continuumlike signal of $\gamma$ rays from the distribution of nonthermal nuclei. In order to characterize this, we first assume that a small fraction of the nuclei undergo nonthermal acceleration. The majority of the nuclei should therefore remain propagating at $\Gamma_\mathrm{bulk}$. We represent this population dichotomy by
\begin{equation}\label{Thermal and Non-thermal Population}
    N_i(\Gamma) =
    \begin{cases}
        f \: N_i, \quad \hfill\Gamma=\Gamma_\mathrm{bulk}, \\
        (1-f) \int^{\Gamma_\mathrm{max}}_{\Gamma_\mathrm{bulk}} \frac{dN_i}{d\Gamma} d\Gamma, \quad \Gamma_\mathrm{bulk} < \Gamma \leq \Gamma_\mathrm{max},
    \end{cases}
\end{equation}
\noindent where $f$ is the fraction that remains thermal, $\Gamma_\mathrm{max}$ is found from the $E_\mathrm{max}$ of the model considered (see Table \ref{BP Models Table}), and $\frac{dN_i}{d\Gamma}$ is the energy distribution of the nonthermal nuclei. These maximum energies are calculated by comparing acceleration timescales to energy loss timescales and are performed in Ref.~\cite{Ekanger:2022tia} (see also Ref.~\cite{Bhattacharya:2021cjc}). This calculation assumes that the average nucleus has the mass of iron ($A=56$) and that magnetic reconnection accelerates nuclei to ultrahigh energies. Throughout the rest of this work, we assume that $f=0.9$, so only 10\% of the nuclei undergo nonthermal acceleration (see also Refs.~\cite{Blasi:2005pb,BLASI_2011,Marti-Devesa:2024hic}). The energy distribution for the nonthermal nuclei is not clear, but as motivated by Ref.~\cite{Murase:2008grb}, we adopt a power law with spectral index $-2$ and include an exponential cutoff at $\Gamma_\mathrm{max}$, i.e.,
\begin{equation}
    \frac{dN}{d\Gamma} \propto \Gamma^{-2} \: \mathrm{Exp}(-\Gamma / \Gamma_\mathrm{max}).
\end{equation}

\begin{figure*}[t]
    \centering
    \includegraphics[width=0.45\linewidth]{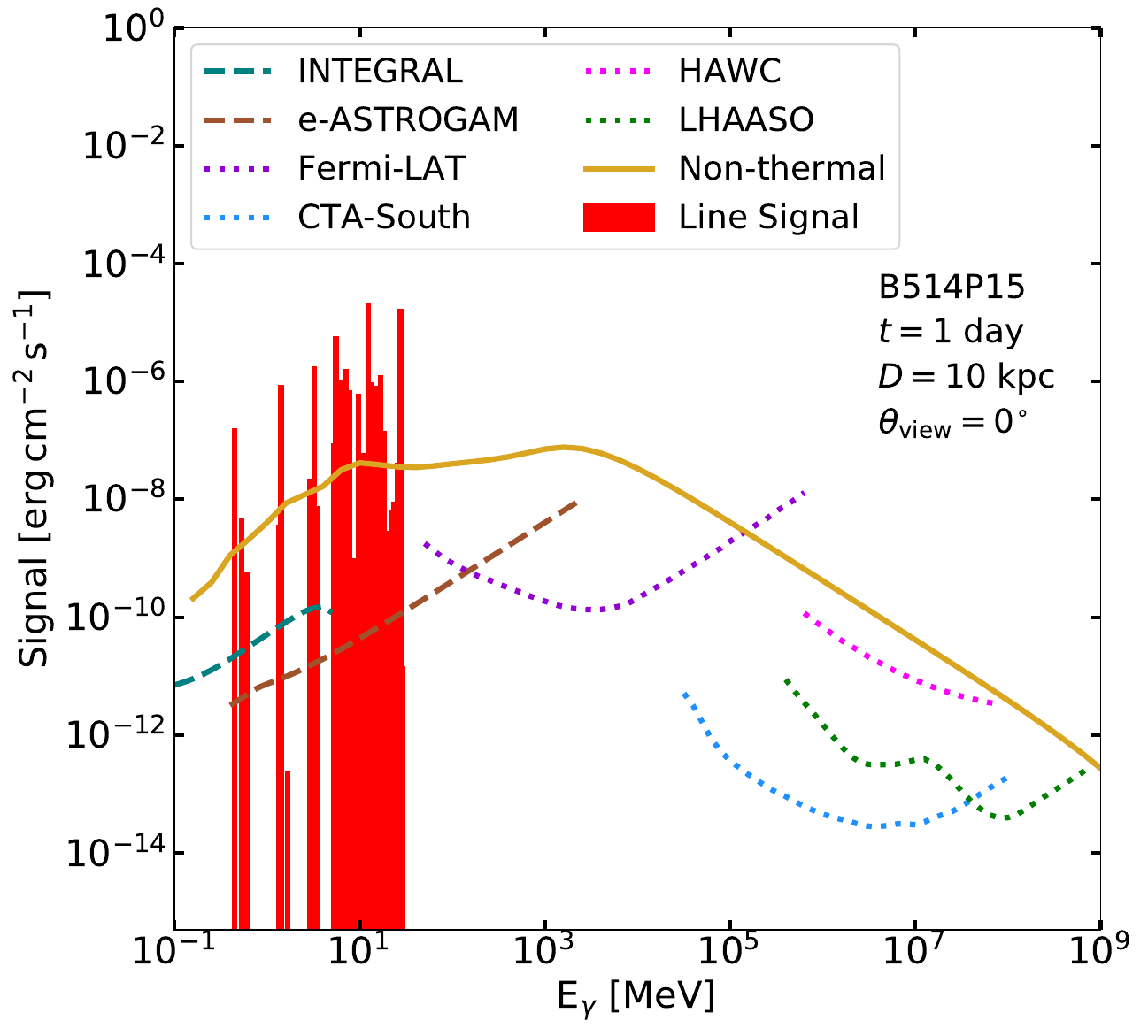}
    \includegraphics[width=0.45\linewidth]{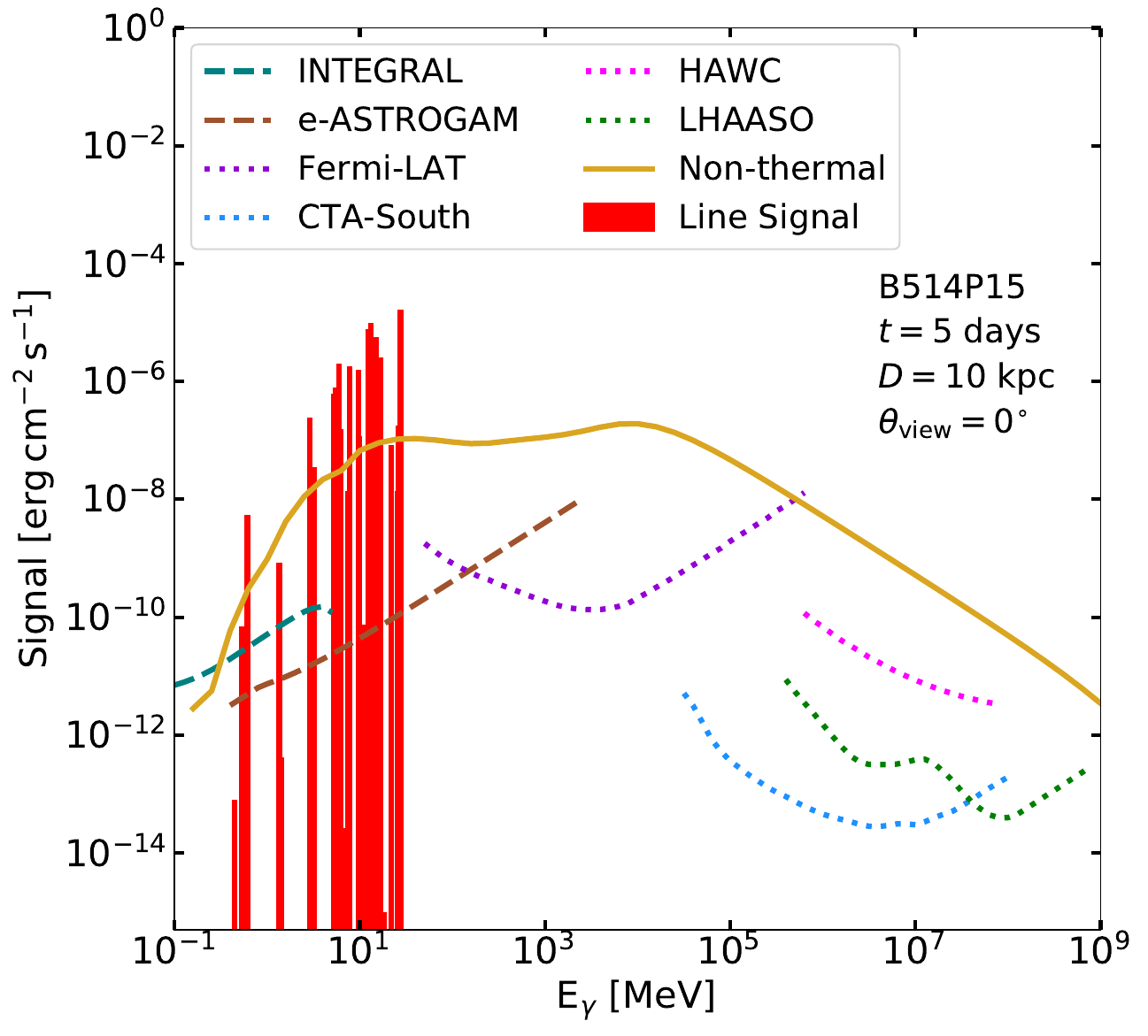}
    \\
    \includegraphics[width=0.45\linewidth]{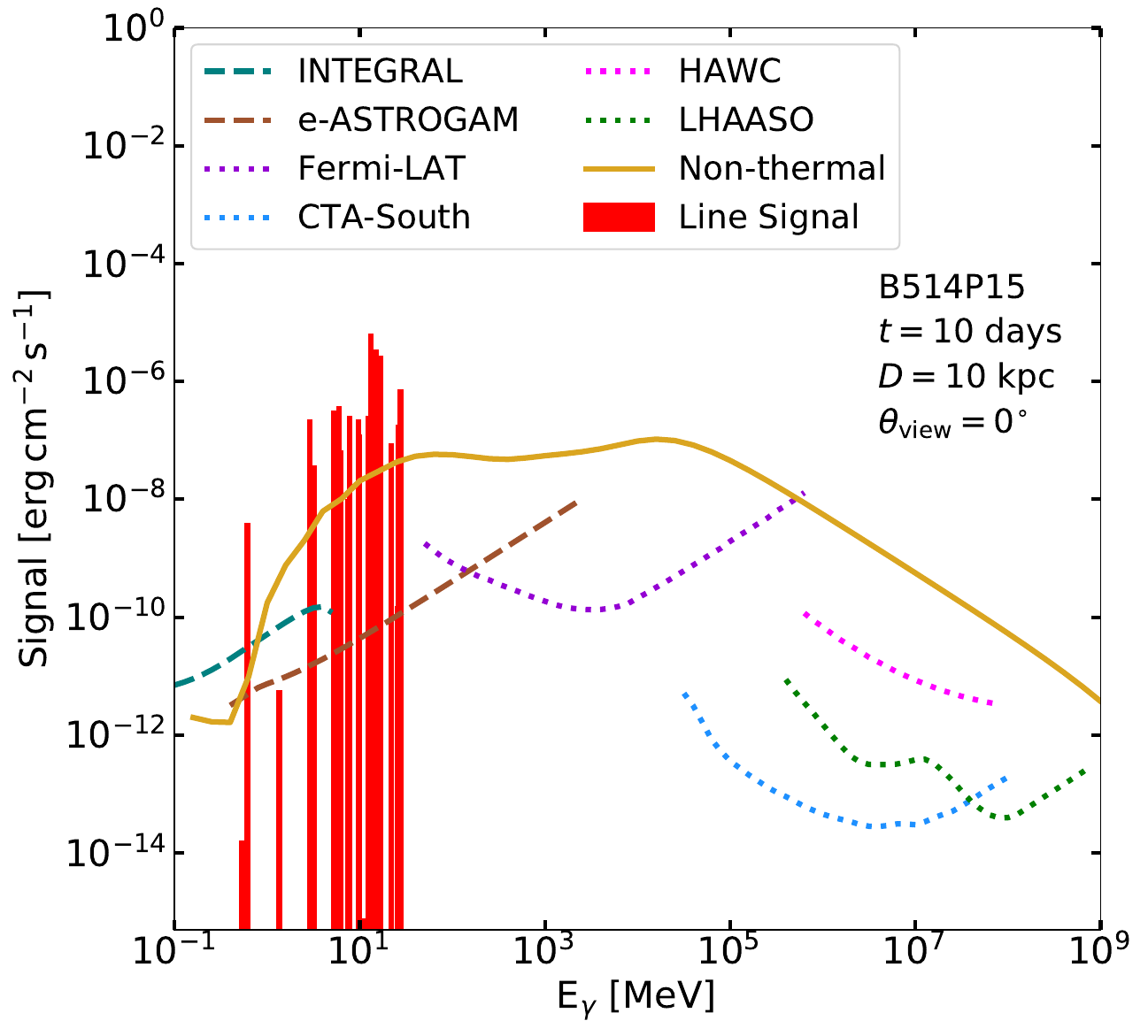}
    \includegraphics[width=0.45\linewidth]{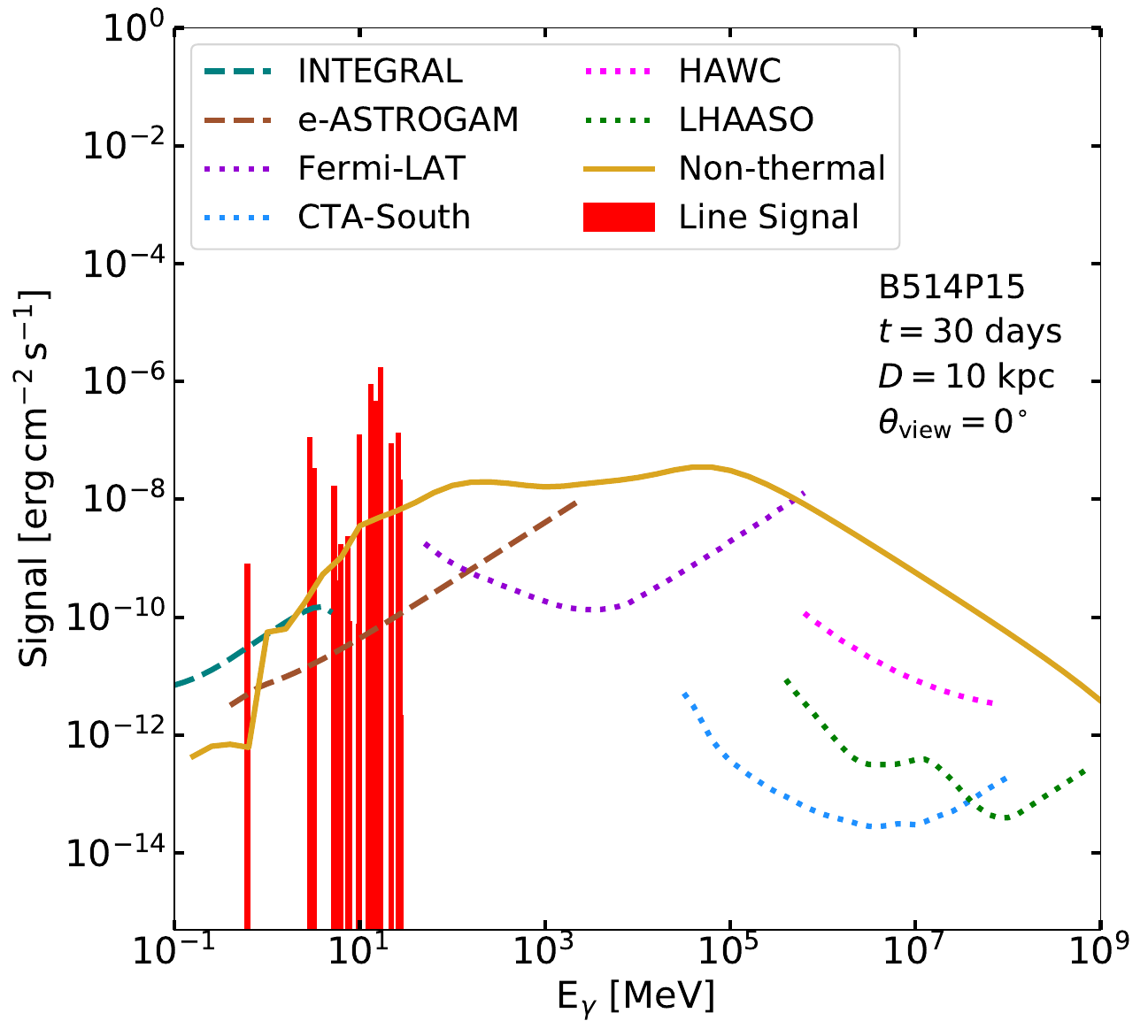}
    \caption{Same as Fig.~\ref{Thermal Line Emission Over Time}, but now includes a nonzero fraction (10\%) of the total population that undergoes acceleration to nonthermal energies (yellow). Red shows the thermal nuclei line signal from Fig.~\ref{Thermal Line Emission Over Time} (red), but rescaled by 0.9. Dashed (dotted) lines show detector point-source line (continuum) sensitivities assuming an observation time of $10^6$\:s.}
    \label{Non-thermal Continuum Emission Over Time}
\end{figure*}

\begin{figure*}[t]
    \centering
    \includegraphics[width=0.45\linewidth]{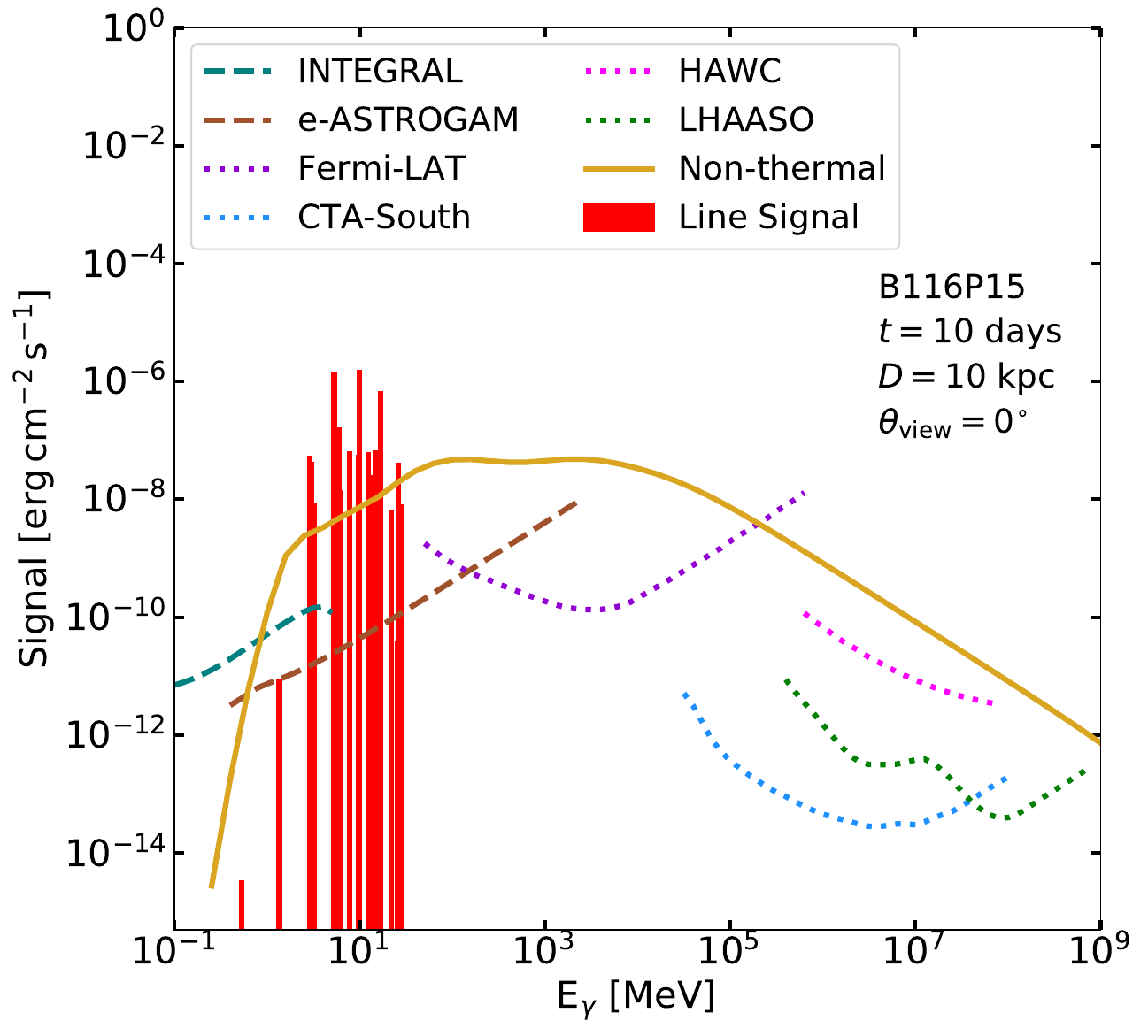}
    \includegraphics[width=0.45\linewidth]{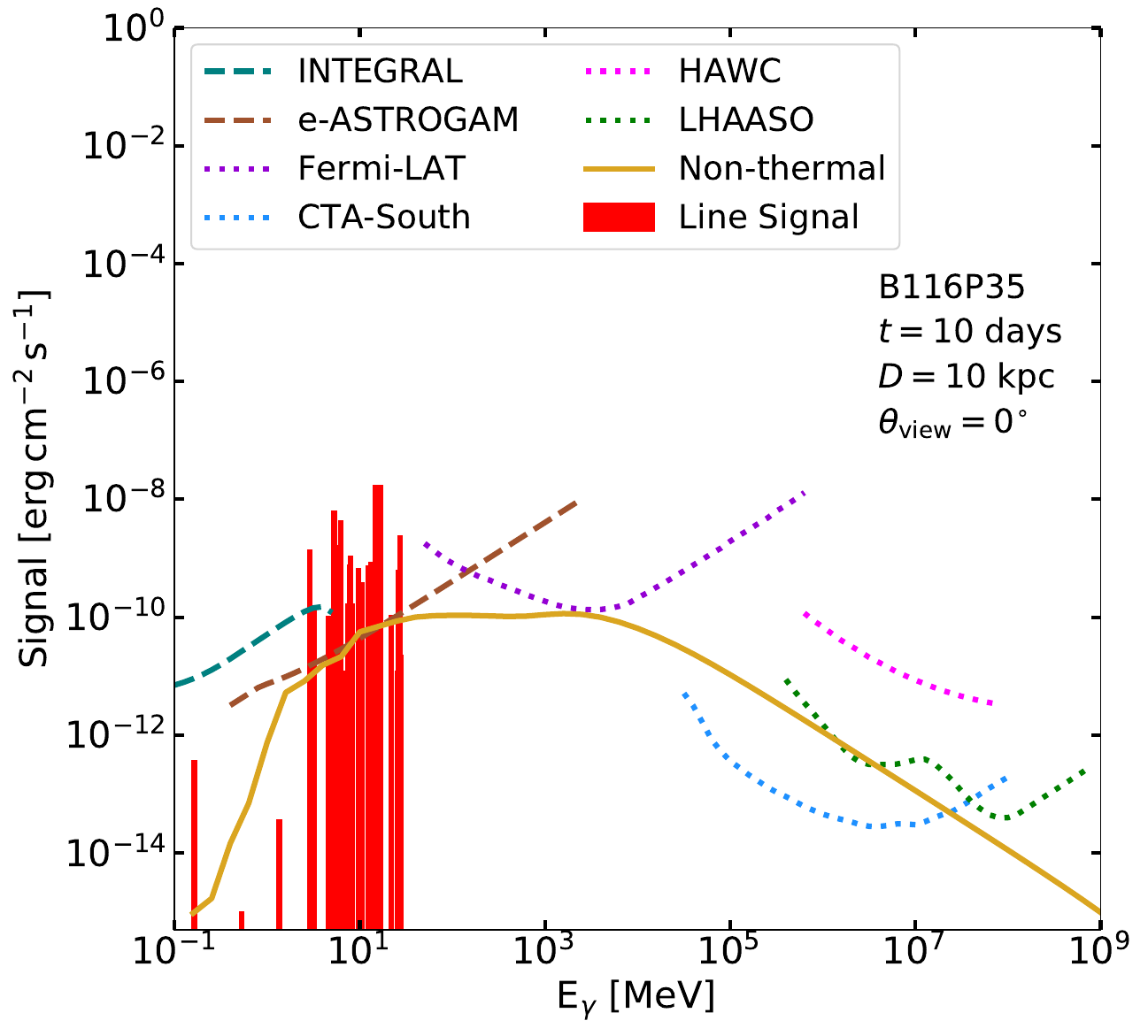}
    \\
    \includegraphics[width=0.45\linewidth]{Continuum_B514P15_Day10.pdf}
    \includegraphics[width=0.45\linewidth]{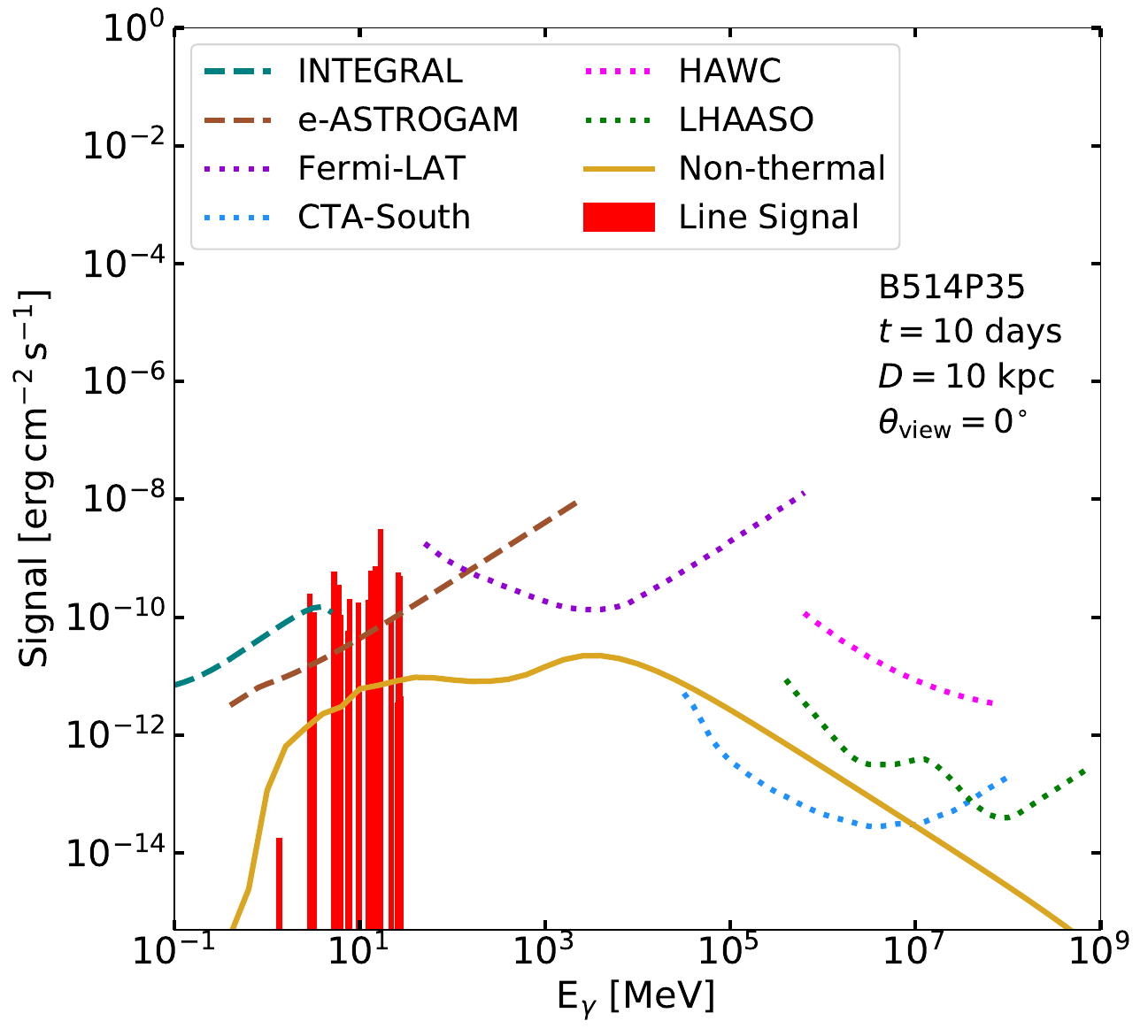}
    \caption{Same as Fig.~\ref{Differing Line Emission t=10 days}, but now includes a nonzero fraction (10\%) of the total population that undergoes acceleration to nonthermal energies (yellow). Red shows the thermal nuclei line signal from Fig.~\ref{Thermal Line Emission Over Time} (red), but rescaled by 0.9. The snapshot is shown at $t=10$\:days, when all models should be optically thin. Dashed (dotted) lines show detector point-source line (continuum) sensitivities assuming an observation time of $10^6$\:s.}
    \label{Differing Continuum Emission t=10 days}
\end{figure*}

Now carrying out the same calculations we did for the thermal signal, we predict the signal from this scenario. As the continuum signal from the nonthermal population will cover a wide range of energies, we keep it unbinned as the energy resolutions of detectors varies from MeV to higher energies. We plot this scenario in Figs.~\ref{Non-thermal Continuum Emission Over Time} and \ref{Differing Continuum Emission t=10 days}. We see that the signal is quite strong across all energies. Here, there is an increase in the luminosity due to the beaming, but this only scales with $\Gamma_\mathrm{bulk}$ and not the individual $\Gamma$'s for higher acceleration. The flatness of the spectrum from $\sim10^2$--$10^5$\:MeV comes from the specific isotopic composition of the jet. The nonthermal continuum slowly dies off only at lower energies over these timescales as the signals come from the nuclei that are accelerated just above $\Gamma_\mathrm{bulk}$, so they do not have their half-lives significantly time dilated. As the signal gets to higher energies, it is near constant over these timescales as the half-lives get boosted to values larger than the observation time after nucleosynthesis.

We see that the nonthermal signal is subdominant to the lines in the 0.1--20\:MeV region. This allows for estimation of the isotopic abundances using the lines as they are not overwhelmed by the nonthermal signal. Also, all of the models produce a detectable nonthermal continuum at \textit{CTA-South} \cite{CTAConsortium:2017dvg} when placed at 10\:kpc for $\sim$10\:days. The models with faster rotations can be seen at even higher energy detectors like \textit{LHAASO} \cite{DiSciascio:2016rgi,LHAASO:2019qtb} and \textit{HAWC} \cite{HAWC:2017osk,HAWC:2021ubt} for on-axis beams. As we are uncertain about the fraction of nuclei that can undergo nonthermal acceleration, we do not calculate the detection horizon or critical angle at 10\:kpc as we did for the pure bulk motion signal.

\subsection{Other messengers from nonthermal acceleration}

If efficient nonthermal particle acceleration occurs in the jets we model, then there could be additional $\gamma$-ray signals, e.g., from neutral pion decays. If nuclei are accelerated above the pion production threshold [$\mathcal{O}(100)$\:MeV], depending on the interaction, nuclei or nucleons in the jet can produce pions through interactions with photons or other nucleon targets. Neutral pions will then decay into two $\gamma$ rays with half the pion rest mass, $E_{\gamma}=m_{\pi}/2\sim70$\:MeV \cite{Fermi-LAT:2013iui,Yang:2018dsi} in the frame of the pion. This results in a $\gamma$-ray number spectrum that is symmetric around $E_{\gamma}$ known as the ``pion bump.'' Because the neutral pion decay half-life is incredibly short ($\sim8\times10^{-17}$\:s \cite{PrimEx-II:2020jwd}), this pion bump is present for as long as there is particle acceleration and sufficient interacting targets. 

We estimate the time frame for when there could be efficient photopion production with nonthermal protons. In order to calculate this, we need to know the photon temperature evolution inside the jet. We get the temperature evolution from the kinetic energy density of the jet, letting it be fully supplied by the radiation, i.e. $u\propto T^4$. We then find the mean-free path for photopion energy loss using the inverse timescale of Ref.~\cite{HighEnergyRad}. We find that initially, the jet is around $10^8$\:K, resulting in a minimum proton Lorentz factor of $\Gamma\sim10^4$ and a mean-free path of $\sim700$\:cm. However, the temperature quickly drops to $\sim10^6$\:K in a day which increases the minimum Lorentz factor and mean free path to $\Gamma\sim10^6$ and $7\times10^8$\:cm, respectively. The jet then slowly cools to several $10^5$\:K over tens of days. This means that early on, photopion production should occur for above threshold protons anywhere in the jet. At later times, if particle acceleration occurs near the leading edge of the jet, then protons should be able to escape before undergoing photopion production with the thermal photons of the jet. Therefore, the photopion signal may only last for days if the particle acceleration occurs near the leading edge of the jet. If this is not the case, then a photopion signal could be produced for tens of days.

Particle acceleration can also lead to the production of charged pions which decay into muon and electron neutrinos. This scenario has been investigated in PM-driven outflows (see, e.g., Refs.~\cite{Bhattacharya:2022btx,Carpio:2023wyt}) and produces much higher energy neutrinos than the thermal $\mathcal{O}(10)$\:MeV neutrinos from CCSNe. Particle acceleration can also produce (ultra) high energy cosmic rays. These charged particles are not coincident in time with these other messengers, however, because their travel times are delayed by intervening magnetic fields.

\section{Discussion}\label{Discussion}

The modeling we have carried out in this paper suggests that, with accurate and precise measurements of the $\gamma$-ray light curve and spectrum, it could be possible to estimate properties of the PM. The difficulty is dealing with the numerous degeneracies. For example, based only on the line emission light curves, the degeneracy between the initial spin period and the magnetic dipole field strength cannot be broken, but combining multiple measurement types can help. Observation of the line signal in combination with the light curve can open the possibility to estimate the mass of the jet and the relative abundances of the unstable nuclei. In this particular example, the mass of the jet is directly proportional to the overall luminosity, and the relative strength of the lines to each other determines the relative abundances. 

If identification of individual lines is made (without blending), then the favorable isotopes laid out in Table \ref{Important Isotopes} can help to constrain $P_i$. Then, with $P_i$ estimated, one can work out the allowed values for $B_\mathrm{dip}$ from the luminosity. There are still other uncertainties with the luminosity that must also be considered, e.g., distance to the source, jet opening angle, viewing angle, and bulk Lorentz factor must be considered as well. 

A null detection from a galactic CCSNe would be harder to interpret, given the large number of different possible reasons, although combining other observables could help. For example, a jet may have been formed but may be off axis ($\theta_{\rm view}\sim90\%$), so the signal may be otherwise faint as the relativistic beaming decreases signals that are significantly off axis; or, the $\gamma$-ray luminosity may simply be too faint due to low ejecta masses, while the CCSNe is still detectable with optical telescopes. Additionally, the ejecta may be too opaque to $\gamma$ rays. Here, there is some uncertainty related to the opacity of ejecta postnucleosynthesis, which is likely a temperature and time dependent quantity, and could be higher than the $0.1$\:g$^{-1}$\:cm$^2$ assumed here (see Refs.~\cite{Tanaka:2019iqp,Raaijmakers:2021slr,Lund:2022bsr}). If the opacity is $1$\:g$^{-1}\:$cm$^2$, for example, the lines are largely unchanged at ten days after nucleosynthesis. Finally, in cases where $Y_e\gtrsim0.5$, only low mass number isotope decay lines would be seen since very few heavy elements are synthesized above the iron-group elements. Thus, the interpretation would depend on whether optical and other observables reveal about, e.g., the progenitor, ejecta, and jets. If nonthermal acceleration occurs in the jet, the effects of relativistic beaming make it difficult to be able to determine the isotopic composition, as the nonthermal continuum signal can be as strong as the line signal in certain scenarios. 

The $\beta^\pm$ decays we use to model the $\gamma$-ray signal also produce neutrinos, which are not monochromatic and depend on the $Q$ value of the decay and energy of the $\beta^\pm$ particle emitted in the decay. These neutrinos should also have their energies boosted due to relativistic effects by a factor of $\delta$. With typical $Q$ values of $0.1$--$10$\:MeV, this means that some of these neutrinos can be detectable by neutrino experiments.  We can perform a simple calculation for the maximum possible neutrino flux at the source, assuming all nuclei decay at the same time. For the most massive jet (B514P15), there are $\sim 10^{51}$ unstable atoms (coming from the abundances and the mass of the jet) which results in $\sim10^{51}$ neutrinos from $\beta^{\pm}$ decays. A typical CCSNe releases a burst of $\sim10^{58}$ thermal neutrinos of energy $\sim10$\:MeV which would result in $\mathcal{O}(10^{4-5})$ detections at Hyper-Kamiokande \cite{hyperkamiokande} if it occurs at $10$\:kpc. With relativistic beaming on axis, $\sim10^{51}$ neutrinos would result in $\mathcal{O}(10^{1-2}$) detections at Hyper-Kamiokande over a time window of days to tens of days (assuming the neutrino emission occurs over the same time frame as the gamma-ray emission). Other CCSN neutrino experiments, like Super-Kamiokande, JUNO, IceCube, and DUNE, will also be able to detect this flux for closer CCSNe. This additional neutrino signal will likely be differentiable from the typical 10\:s neutrino emission because the jet-based neutrino emission occurs over much longer time frames, on the scale of days.

Our framework could be applied to other transients that produce heavy elements like compact object mergers. However, as the velocity of the mass ejected may be subrelativistic \cite{Villar:2017wcc,Chen:2021tob} it is unlikely that any $\gamma$-ray signal from decays is boosted and/or beamed significantly. However, it is interesting to point out that there are some overlapping nuclei decay lines of interest between these scenarios. The authors of Ref.~\cite{Chen:2021tob}, for example, identify an important $^{72}$Zn line for their $Y_e>0.3$ case---in common with our B514P15 and B116P15 models (see also Ref.~\cite{An:2023edd}). Mergers could produce more decay lines and also eject relatively more mass \cite{Amend:2024sdm}, but they occur much less frequently than supernovae locally. The volumetric occurrence rate for binary neutron star mergers is $\sim10^2$\:Gpc$^{-3}$\:yr$^{-1}$ while the rate for magnetorotational supernovae is $\sim1\%\:\Dot{n}_{\rm CC}\sim10^3$\:Gpc$^{-3}$yr$^{-1}$ \cite{Nishimura:2015nca,Kashiyama:2015eua,Vlasov_2017,Halevi:2018vgp} where $\Dot{n}_{\rm CC}\sim10^5$\:Gpc$^{-3}$yr$^{-1}$ is the volumetric rate of typical CCSNe \cite{Taylor:2014rlo}.

There are various uncertainties in the modeling we carry out in this work. The nucleosynthesis yields depend sensitively on the choice of $Y_e$ and, while we choose a moderately neutron-rich fraction of $Y_e=0.45$, it could be lesser or greater than this in PM outflows. We choose this in order to guarantee some nonzero population of unstable nuclei, but it is also motivated by both analytical and simulation works \cite{Vlasov_2017,Halevi:2018vgp,Reichert:2024vyd}. There are also uncertainties in the physical model of the jet, such as the opening angle of the jet (e.g., for long GRBs the uncertainty can be tens of degrees \cite{Fong_2012}), which affects the optical depth. The surrounding CSM is also uncertain, and heavily depends on if there is a preferred progenitor type for PM formation. Additionally, if there is a binary companion, there may be a dense common envelope formed between the stars \cite{Hurley:2002rf,Ivanova_2013}, which itself would be able to cause the afterglow phase to begin earlier on than our current CSM treatment.

In our modeling, we only consider the $\gamma$ rays from $\beta^{\pm}$ decays, but $\gamma$ rays from isomeric transitions out of metastable nuclear states may also be important signals. Although we exclude them in this work, some long-lived transitions may give rise to a signal comparable to that seen from $\beta^{\pm}$ decays, like $^{111}$Pd which is synthesized in large abundances in our models. Many isomeric transitions have half-lives on the order of $\mathcal{O}(1)$\:ns, but some could be detectable in the same energy range considered here. Including isomeric transitions may change the way nuclei are synthesized in neutron capture processes, modify the heating rate related to kilonova time evolution, and emit detectable $\gamma$ rays \cite{Kajino:2019abv,Fujimoto:2020aub,Misch:2020gnm,Misch:2020axi,Misch:2024adc,Hamilton:2024tho}, and the extent to which these effects are important are actively being studied. Some recent nuclear reaction networks, like {\tt WinNet}, can take into account isomeric transitions of $^{26}$Al, for example \cite{Reichert:2023xqy}.

\section{Summary}\label{Summary}

In this work, we have modeled $\gamma$-ray signals that come from the decays of heavy nuclei within PM jets. During the core collapse of the progenitor, neutrino-driven outflows are collimated and launched as a relativistic jet that is powered by the rotation and magnetic field of the PM. Nuclei could be synthesized in these jets, potentially up to weak $r$-process elements. As the jet propagates, it expands and carries the nuclei outside the progenitor, and eventually becomes optically thin, allowing $\gamma$ rays from the decays of radioactive isotopes to escape. These $\gamma$-ray lines have their energies and luminosities boosted due to the bulk Lorentz factor of the jet. If nuclei are further subjected to particle acceleration, a nonthermal tail in $\gamma$ rays is also possible. The light curves and spectra of the $\gamma$-ray emission are sensitive to the mass ejected and the isotopic abundances, allowing for dependencies on central engine parameters $P_i$ and $B_\mathrm{dip}$. 

Our modeling uses a simple physical model for the jet and models of the surrounding CSM. The CSM models allow for an estimation of the time at which the jet enters the afterglow phase, which is likely the point at which the signal from the boosted $\gamma$ rays will be overwhelmed by the afterglow signal as the jet decelerates. We primarily focus on on-axis jets, and the signal should greatly vary as the viewing angle becomes more off axis. 

Detection of this signal should be possible for the models we consider with current and future $\gamma$-ray detectors. The detection horizon is most sensitively dependent on the ejecta mass and the viewing angle of the jet. Because of relativistic beaming, in the best case scenario the signal is detectable out to $\sim35$\:Mpc for on-axis jets just for the thermal signal, but is still extragalactic for the less favorable models. For cases in which nonthermal acceleration is possible, the detection horizon should also be extragalactic, but this also highly depends on the amount of nuclei that can accelerate to these higher energies and the population distribution they follow. Off-axis jets, however, are typically limited to be detectable within our galaxy or only in the nearest galaxies to the Milky Way, depending on the viewing angle.

Detection of this signal would help to test the physical origin and mechanisms of the PM jet model. It would serve as a way to confirm abundant nucleosynthesis in relativistic ejecta, motivating the presence of magnetorotational CCSNe. These magnetorotational powered supernovae are thought to potentially power a range of rarer supernovae, and boosted line measurements could help probe their central engine properties. 

\section*{Acknowledgements}
We thank Gonzalo Herrera for helpful discussions. S.~Heston is supported by NSF Grant No.~PHY-2209420, the Julian Schwinger Foundation, and JSPS KAKENHI Grant Number JP23H04899. 
N.~Ekanger is supported by NSF Grant No.~AST1908960.
The work of S.~Horiuchi is supported by the U.S.~Department of Energy Office of Science under award number DE-SC0020262, NSF Grant No.~AST1908960 and No.~PHY-2209420, the Julian Schwinger Foundation, and JSPS KAKENHI Grant Number JP22K03630 and JP23H04899. This work was supported by World Premier International Research Center Initiative (WPI Initiative), MEXT, Japan.

\section*{Data Availability}
The data supporting this study's findings are available within the article.

\bibliography{research.bib}

\end{document}